\documentclass[aps,prl,twocolumn,superscriptaddress,showpacs,preprintnumbers,nofootinbib]{revtex4-2}

\usepackage{epsfig}
\usepackage{multirow}
\usepackage{slashed}
\usepackage{amsmath}
\usepackage{amssymb,amsthm,amscd}
\usepackage{array}
\usepackage{physics}
\usepackage{quantikz}
\usepackage{braket}
\usepackage{graphicx}
\usepackage{subfigure}
\usepackage{algorithm}
\usepackage{ulem}
\usepackage{color}
\usepackage{epstopdf}
\usepackage{orcidlink}
\usepackage[inline]{enumitem}

\newcommand{\im}{\mathrm{i}}
\newcommand{\OO}{\mathcal{O}}
\newcommand{\Hfix}{H_{\rm fix}}
\newcommand{\Uax}{U_{\rm fix}}
\usepackage{xurl}
\usepackage{hyperref}
\hypersetup{colorlinks=true, linkcolor=blue, citecolor=blue, urlcolor=blue,breaklinks=true}

\newcommand{\ad}{\operatorname{ad}}
\newcommand{\V}{\mathcal V}
\newcommand{\Fp}{\mathcal F}
\newcommand{\Amax}{A_{\rm max}}
\newcommand{\Aopmax}{A_{\rm op,max}}
\newcommand{\Pimax}{\Pi_{\rm max}}
\newcommand{\Piopmax}{\Pi_{\rm op,max}}
\newcommand{\Gfp}{\Gamma_{\rm FP}}
\newcommand{\Klat}{\mathcal K_{\rm lat}}
\newcommand{\Hred}{H_{\rm red}}
\newcommand{\cP}{c_{\rm P}}
\newcommand{\cad}{c_{\rm ad}}
\newcommand{\poly}{\operatorname{poly}}
\newcommand{\polylog}{\operatorname{polylog}}
\newcommand{\Ran}{\operatorname{Ran}}
\newcommand{\Png}{P_{\rm ng}}
\newcommand{\ExpPsi}[1]{\left\langle #1\right\rangle_{\Psi}}
\newcommand{\Spos}{\mathcal{S}_{\rm pos}}
\newcommand{\Scol}{\mathcal{S}_{\rm col}}
\newcommand{\Slat}{\mathcal{S}_{\rm lat}}
\newcommand{\yy}[1]{\textcolor{blue}{#1}}
\begin{document}
\preprint{RIKEN-iTHEMS-Report-26}

% 标题与作者信息
\title{Efficient Quantum Simulations of Yang-Mills theory with Maximal-tree Gauge}
\author{Tianyin Li}
\email{tianyin.li@riken.jp}
\affiliation{RIKEN Center for Interdisciplinary Theoretical and Mathematical Sciences (iTHEMS), RIKEN, Wako 351-0198, Japan}

\author{Ying-Ying Li}
\email{liyingying@ihep.ac.cn}
\affiliation{Institute of High Energy Physics, Chinese Academy of Sciences, Beijing 100049, China}

\author{Xiaoyang Wang\orcidlink{0000-0002-2667-1879}}
\email{xiaoyang.wang@riken.jp}
\affiliation{RIKEN Center for Interdisciplinary Theoretical and Mathematical Sciences (iTHEMS), RIKEN, Wako 351-0198, Japan}
\affiliation{RIKEN Center for Computational Science (R-CCS), Kobe 650-0047, Japan}

\author{Hongxi Xing}
\email{hxing@m.scnu.edu.cn}
\affiliation{State Key Laboratory of Nuclear Physics and Technology, Institute of Quantum Matter,
South China Normal University, Guangzhou 510006, China}
\affiliation{Guangdong Basic Research Center of Excellence for Structure and Fundamental Interactions of Matter,
Guangdong Provincial Key Laboratory of Nuclear Science, Guangzhou 510006, China}

\date{\today}

\begin{abstract}
We develop a quantum algorithmic framework for the efficient simulation of Yang--Mills theories, including the $\mathrm{SU}(3)$ gauge theory in Quantum Chromodynamics (QCD). The framework uses maximal-tree gauge in terms of gauge field variables that removes all local gauge redundancies. In the resulting gauge-fixed formulation and digitization in the field-amplitude basis, we show that Hamiltonian time evolution admits an efficient implementation based on quantum singular value transformation (QSVT). We derive upper bounds on the total number of qubits and gate complexity, finding polynomial scaling with the inverse simulation precision $1/\varepsilon_s$, lattice volume $\mathcal{V}$, gauge coupling $g$, and target energy scale $E$. Our results provide a rigorous complexity-theoretic demonstration that non-Abelian Yang--Mills theories can be simulated efficiently on quantum computers, paving the way toward first-principles quantum simulations of non-perturbative QCD dynamics.
\end{abstract}

\maketitle

% --- 正文开始 ---

\vspace{0.5cm}
\noindent{\it Introduction:}
Quantum Chromodynamics (QCD), the fundamental theory of the strong interaction, describes the dynamics of quarks and gluons. Its gauge sector is governed by an $\mathrm{SU}(3)$ non-Abelian Yang--Mills theory. While the QCD Lagrangian follows elegantly from local $\mathrm{SU}(3)$ gauge symmetry, its long-distance dynamics are non-perturbative, making first-principles studies of hadron structure, hadron scattering, and the QCD phase diagram highly challenging. Lattice QCD (LQCD) based on Monte Carlo methods has achieved remarkable success in computing non-perturbative static observables~\cite{Hagler:2009ni,Lin:2017snn,Briceno:2017max,Meyer:2022mix,FlavourLatticeAveragingGroupFLAG:2024oxs}, but suffers from the notorious sign problem in real-time dynamics and at finite fermion density~\cite{deForcrand:2002hgr,Borsanyi:2020fev,Bazavov:2017dus}.

Quantum computing offers a complementary, sign-problem-free approach to real-time and finite density quantum field theory \cite{Bauer:2022hpo, DiMeglio:2023nsa, Fang:2024ple}. In particular, efficient quantum simulation of non-Abelian Yang--Mills (YM) theories is a fundamental step toward first-principles quantum simulation of QCD. Quantum simulation of quantum field theory requires both spatial discretization and truncation of the infinite-dimensional local Hilbert space to a finite-dimensional space~\cite{Byrnes:2005qx,Jordan:2012xnu}. For non-Abelian gauge theories, several approaches have been developed, including truncations in the electric or representation basis~\cite{Byrnes:2005qx,Klco:2020xrg,Ciavarella:2021qnu,Ciavarella:2025bsg,Balaji:2025yua}, large-\(N_c\) expansions~\cite{Ciavarella:2024fzw}, gauge-invariant loop-string-hadron formulations~\cite{Raychowdhury:2020xzx,Raychowdhury:2020,Kadam:2023}, and renormalized dual basis~\cite{Fontana:2024rux,Miranda-Riaza:2025fus}. Field-amplitude digitization has been explored in Coulomb and axial gauges~\cite{Li:2024ide,Yao:2025uxz,Yao:2026rya}, which face, respectively, the Gribov ambiguity and incomplete gauge fixing~\cite{Gribov:1977wm,Vandersickel:2012tz,Christ:1980ku,Chodos:1978pe,Schwinger:1962fg}. Discrete-subgroup formulations provide an alternative finite-dimensional construction for $\mathrm{SU}(2)$ and $\mathrm{SU}(3)$, although systematic control of the continuum limit remains an active area of investigation.~\cite{Lamm:2024jfz,Gustafson:2024,Gustafson:2024vju}.

Among these methods, quantum simulation efficiency of the non-Abelian gauge theories remains lacking because physical states must satisfy Gauss's law, which imposes $(N_c^2-1)\mathcal{V}$ local constraints for $\mathrm{SU}(N_c)$ YM theory on a lattice with $\mathcal{V}$ sites. These constraints reflect the local continuous $\mathrm{SU}(N_c)$ gauge redundancy and introduce substantial complexity in the quantum simulation.

In this work, we show that YM theory can be simulated efficiently on quantum computers. We eliminate the local gauge redundancy by choosing the maximal-tree gauge. While maximal-tree gauge fixing has been explored at the level of link variables~\cite{DAndrea:2023qnr,Carena:2024dzu}, we formulate it directly in terms of the gauge field variables. The resulting gauge fixing conditions are $A^a_3(n_1,n_2,n_3) = 0,
A^a_2(n_1,n_2,n_3=0) = 0,
A^a_1(n_1,n_2=0,n_3=0) = 0$.
We then perform field-amplitude digitization, truncating the physical gauge field \(A^a_i\) to a finite-dimensional set. In this formulation, the temporal component \(A^a_0\) is determined by Gauss's law and can be efficiently implemented using quantum singular value transformation (QSVT)~\cite{PhysRevLett.118.010501,Gilyen:2018khw,PRXQuantum.2.040203,chakraborty2025quantumsingularvaluetransformation}, enabling a rigorous quantification of the quantum resources required for simulating YM theory. We provide the first rigorous derivation of the qubit cost $N_G$ required to encode the gauge field degrees of freedom:
$$N_G \sim \mathcal{V} \log_2\left(\frac{\mathcal{V}^{3/2} E}{\varepsilon_s}\right)\,,$$
as well as the gate complexity $\mathcal{M}_U$ for simulating its time-evolution operator:
$$\mathcal{M}_U \sim O\left(\frac{t g^2 \mathcal{V}^{20/3} E^2}{\varepsilon_s^2}\right)\,.$$
Both costs grow at most polynomially with respect to the system size $\mathcal{V}$, energy scale $E$, YM coupling $g$, and the inverse simulation precision $\varepsilon^{-1}_s$. Our results thus offer a resource-efficient baseline for quantum simulations of YM theory.

\vspace{0.5cm}
\noindent{\it Hamiltonian of Yang-Mills theory and gauge fixing:} 
The Hamiltonian of YM theory without any gauge fixing is
\begin{align}\label{eq:comtiham}
    H_{\rm YM}
     &= \int d^3x \left[-\frac{1}{2} \Pi^a_i \Pi^{ia} + \frac{1}{4} F^a_{ij} F^{ija} - A_{0}^b \mathcal{G}^b\right]\,,
\end{align}
where $i=1,2,3$ denote the $x$, $y$, and $z$ directions, respectively, and $\mathcal{G}^b = D^{ab}_i \Pi^{ia}$ is Gauss's law operator. On the lattice, there are $\mathcal{{V}} = {L}^3$ lattice sites and the lattice spacing is $a_s$. Without loss of generality, we set $a_s=1$. 

\begin{figure}[t]
  \centering
     \includegraphics[width=0.8\columnwidth]{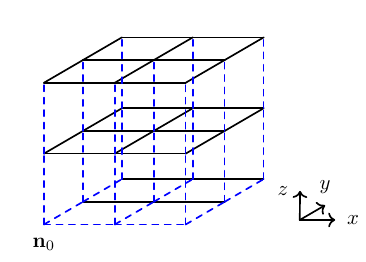}
  \caption{Maximal tree (dashed lattice links) for a cubic lattice with open boundary conditions, where $\mathbf{n}_0$ represents the origin.}
  \label{fig:gauge-fixing}
\end{figure}

We consider a lattice with open boundary conditions, for which the maximal-tree gauge can be chosen as shown in Fig.~\ref{fig:gauge-fixing}: the link variables along the $z$ direction are set to the identity throughout the lattice, those along the $y$ direction are
set to the identity on the $n_3=0$ boundary, and those along the $x$
direction are set to the identity on the $n_2=n_3=0$ boundary. With all local gauge redundancies removed, these conditions can be expressed directly in terms of the gauge field variables as:
\begin{align}\label{eq:gaufix}
&A^a_3(n) = 0\,,\nonumber\\
    &A^a_2(n_{12},n_3 = 0) = 0\,,\nonumber\\
    &A^a_1(n_1,n_{23}=0) = 0\,,
\end{align}
where we have defined $n_{ij}=(n_i,n_j)$, and $n=(n_1,n_2,n_3)$. The gauge transformation at the origin $\mathbf{n}_0$ remains unfixed, corresponding to a residual global gauge transformation.

\begin{table*}[htbp]
\begin{tabular}{|c|c|c|c|}
\hline
Vector index & primary constraints & secondary constraints & physical variables 
\\ \hline
$\mu = 0$ & $\Pi^a_0(n) = 0$ & $A^a_0(n) = -\sum_m (K^{-1})_{n,m}\mathcal{Q}^a(m) $ &  
\\ \hline
$\mu = 1$    & $A_1^a(n_1, n_{23}=0)= 0$  & $\Pi_1^a(n_1, n_{23}=0) = \partial^R_1 A_0^a(n_1, n_{23}=0) $ & $A_1^a(n_1,n_{23} \not= 0)$, $\Pi_1^a(n_1,n_{23} \not= 0)$  
\\ \hline
$\mu = 2$ & $A_2^a(n_{12},n_3 = 0) = 0$ & $\Pi_2^a(n_{12},n_3 = 0) = \partial_2^R A_0^a (n_{12},n_3 = 0) $ & $A_2^a(n_{12},n_3 \not= 0)$, $\Pi_2^a(n_{12},n_3 \not= 0)$ 
\\ \hline
$\mu = 3$ & $A_3^a(n) = 0$ & $\Pi_3^a(n) =\partial^R_3 A_0^a(n)$ & 
\\ \hline
\end{tabular}
\caption{Summary of the gauge-fixing conditions, the resulting secondary constraints, and the remaining physical variables. Here, $\mu$ runs over the temporal and spatial directions. The secondary constraints are obtained by calculating the Poisson bracket between the primary constraints and the Hamiltonian.}
\label{tab:variables}
\end{table*}
Together with the constraint $\Pi^a_0(n)=0$ on the canonical momentum conjugate to $A^a_0$, we list all primary constraints and their corresponding secondary constraints in Tab.~\ref{tab:variables}, with the latter obtained from the Poisson brackets of the primary constraints with the Hamiltonian. The secondary constraints can be solved for the dependent field degrees of freedom, leaving the two sets of physical variables listed in the last column of Tab.~\ref{tab:variables}. In particular, the temporal gauge field $A_0$ is determined by the Gauss's law constraint as
\begin{align}
    KA^a_0 + \mathcal{Q}^a = 0,
    \label{eq:v4-FP-eq}
\end{align}
where $\mathcal{Q}^a$ is the Gauss's law operator in terms of only physical variables.  $K$ is a matrix that acts on position space given by
\begin{align}\label{eq:v4-Kdef}
    &K
    =
    -\partial_3^2
    -
    \delta_{n_3,0}\partial_2^2
    -
    \delta_{n_2,0} \delta_{n_3,0} \partial_1^2\,,
\end{align}
where $\partial_i^2$ are the second-order derivatives on the lattice. The Kronecker deltas $\delta_{n_i,0}$ indicate that the corresponding second-order derivatives are evaluated only at the specified boundaries.
With Neumann boundary conditions (NBCs), the operator $K$ has a constant
zero mode corresponding to the residual global gauge transformation left
unfixed by the maximal-tree gauge at $\mathbf{n}_0$. We restrict to the physical
global charge-singlet subspace, for which the global Gauss-law constraint
\begin{equation}
    \sum_n \mathcal{Q}^a(n)=0
\end{equation}
is satisfied. On this physical subspace, with the constant zero mode excluded, $K$ is
positive definite and therefore invertible. Gauss's law can therefore be solved as $A^a_0 = -K^{-1} \mathcal{Q}^a$. We use one-sided finite differences (see endnote) in deriving Eq.~(\ref{eq:v4-Kdef}), which is essential for ensuring the positive definiteness of $K$.

Promoting these conditions to the operator level, we construct the Hamiltonian for YM dynamics entirely in terms of the two sets of physical operators, which satisfy the following commutation relations:
\begin{align}\label{eq:comAPI}
    &[{A}^a_1(n_1,n_{23}\not=0),{\Pi}_1^b(m_1,m_{23}\not=0)] = -{\rm i} \delta^{ab} \delta_{n,m}\,, \nonumber\\
    &[{A}^a_2(n_{12},n_3\not=0),{\Pi}^b_2(m_{12},m_3\not=0)] = -{\rm i} \delta^{ab} \delta_{n,m}\,.
\end{align}
In the following, we use the index $I$ with $I = 1, 2$ to denote the two sets of physical variables, especially ${\Pi}^a_{I, \rm Phys}$ with $\Pi^a_{1,{\rm Phys}} = \Pi^a_1(n_1,n_{23} \neq 0)$ and $\Pi^a_{2, {\rm Phys}} = \Pi^a_2(n_{12},n_3 \neq 0)$, and $A^a_{I, \rm Phys}$ with $A^a_{1,{\rm Phys}} = A^a_1(n_1,n_{23} \neq 0)$ and $A^a_{2, {\rm Phys}} = A^a_2(n_{12},n_3 \neq 0)$.
In the gauge defined by Eq. \eqref{eq:gaufix}, the YM Hamiltonian is
\begin{align}\label{eq:hax1}
    \Hfix =& H_{\Pi} + H_{B} + H_{V}\,.
\end{align}
Here, $H_\Pi= \frac{1}{2}\sum_{n} ({\Pi}^a_{I, \rm Phys})^2$ is the electric energy, while $H_B = \frac{1}{2} \sum_{n} \sum_{i>j} (F^a_{ij})^2$ is the magnetic energy. $H_{V}$ is the non-local interaction that comes from the gauge fixing
\begin{align}
    H_{V} = \frac{1}{2} \sum_{n,m} [\mathcal{Q}^a(n)]^\dagger (K^{-1})_{n,m} \mathcal{Q}^a(m)\,.
\end{align}

We would like to point out that the discretization error of the pure-gauge Hamiltonian in $\Hfix$ is $O(a_s)$. This is because $H_V$ contains terms that are linear in first derivatives. Replacing these derivatives by one-sided finite differences on the lattice introduces an $O(a_s)$ discretization error.

\vspace{0.5cm}
\noindent{\it Digitization of gauge fields and qubit cost: }For simplicity, we use a collective index $\xi$ to denote all indices of the gauge fields, including the color index, the index labeling the two physical operators, and the spatial-coordinate index specifying the lattice site. We digitize the gauge field in the field-amplitude basis in which the gauge-field operator ${A}_\xi$ is diagonal,
\begin{align}
    {A}_\xi \ket{A}_\xi = A\ket{A}_\xi\,,
\end{align}
where $-A_{\rm max} \leq A = -A_{\rm max}+\lambda \delta_A \leq A_{\rm max}$, with $\lambda = 0,1,2,...,2^\mathcal{K}-1$, and $\delta_A = 2A_{\rm max}/(2^\mathcal{K}-1)$ is the spacing between adjacent eigenvalues of ${A}_\xi$. On the other hand, the conjugate-momentum basis of ${\Pi}_\xi$, denoted by
$\ket{\Pi}_\xi$, is obtained from the $\ket{A}_\xi$ basis by a discrete
quantum Fourier transformation. The truncation scale of ${\Pi}_\xi$, denoted by $\Pi_{\rm max}$, is of $O(1/\delta_A)$, with a spacing $\delta_\Pi \sim 1/A_{\rm max}$.

We characterize the truncation error $\varepsilon_s$ of the wavefunction $\ket{\Psi}$ to estimate the required number of qubits. Generally, a wavefunction $\ket{\Psi}$ satisfies
\begin{align}
    \sum_A|\langle \Phi_A| \Psi \rangle|^2 = \begin{cases}
        1-\varepsilon_s \,, \ {\rm if}\ |A_{\xi = 1}|,...,|A_{\xi = \xi_{\rm max}}| \leq A_{\max}\,,\\
        \varepsilon_s\,, \ {\rm others}\,,
    \end{cases}
\end{align}
where $\xi_{\rm max}$ is total number of collective indexes, and $\ket{\Phi_A} \equiv \otimes_{\xi} \ket{A}_{\xi}$. There is also a similar truncation error of the wavefunction in the $\ket{\Phi_{\Pi}} = \otimes_\xi \ket{\Pi}_\xi$ basis. The truncation scale $A_{\rm max}$ and $\Pi_{\rm max}$ should be determined by the desired $\varepsilon_s$, 
which is given by~\cite{Jordan:2012xnu}
\begin{align}\label{eq:APim}
    A_{\rm max}
    &\sim O\left(
    \sqrt{\frac{2(N_c^2-1)\mathcal{V}}{\varepsilon_s}
    \braket{A_\xi^2}}\right), \nonumber\\
    \Pi_{\rm max}
    &\sim O\left(
    \sqrt{\frac{2(N_c^2-1)\mathcal{V}}{\varepsilon_s}
    \braket{\Pi_\xi^2}}\right).
\end{align}
Throughout this section, expectation values denoted by $\langle\cdots\rangle$
are understood to be taken with respect to the wavefunction $\ket{\Psi}$. As $\braket{A_\xi^2}$ and $\braket{\Pi_\xi^2}$ contribute to the energy $E=\braket{\Hfix}$ of the wavefunction $\ket{\Psi}$, they can be bounded by $E$.

We begin with
\begin{align}
    E = \braket{H_\Pi} + \braket{H_B} + \braket{H_V}.
\end{align}
Since $K$ is positive definite, we have $\braket{H_V}\geq 0$. In addition, we can prove that $\braket{H_B} \geq \frac{1}{4\mathcal{V}}\sum_\xi \braket{A_\xi^2}$ as follows.

Consider first the $F^a_{32}(n)$ term in $H_B$, with $F^a_{32}=\partial^R_3 A^a_2$. Using the boundary condition $A^a_2(n_{12},0)=0$, we have
\begin{align}
    A^a_2(n_{12},n_3)
    =\sum_{m_3=0}^{n_3-1}F^a_{32}(n_{12},m_3).
\end{align}
The Cauchy--Schwarz inequality then gives
\begin{align}\label{eq:AF32}
    \left\langle A^a_2(n)^2\right\rangle
    \leq
    L\sum_{m_3=0}^{L-1}
    \left\langle F^a_{32}(n_{12},m_3)^2\right\rangle,
\end{align}
and hence
\begin{align}
    \sum_n\left\langle A^a_2(n)^2\right\rangle
    \leq
    L^2\sum_n\left\langle F^a_{32}(n)^2\right\rangle.
\end{align}
Similarly,
\begin{align}
    \sum_n\left\langle A^a_1(n)^2\right\rangle
    \leq
    2L^3\sum_n
    \left\langle
    F^a_{31}(n)^2+F^a_{21}(n)^2
    \right\rangle.
\end{align}
Combining these inequalities, we obtain
\begin{align}\label{eq:HBbound}
    \left\langle H_B\right\rangle
    \geq
    \frac{1}{4\mathcal{V}}
    \sum_{n}\left\langle A^a_{I, \rm Phys}(n)^2\right\rangle
    =
    \frac{1}{4\mathcal{V}}
    \sum_\xi\left\langle A_\xi^2\right\rangle.
\end{align}
With the above bounds on $\braket{H_B}$ and $\braket{H_V}$, we have
\begin{align}\label{eq:EAPi}
    E \geq \braket{H_\Pi} + \braket{H_B}
    \geq \frac{1}{2}\sum_\xi\braket{\Pi_\xi^2}
    + \frac{1}{4\mathcal{V}}\sum_\xi\braket{A_\xi^2}.
\end{align}
Since each term on the right-hand side of Eq.~\eqref{eq:EAPi} is positive semidefinite, we obtain
\begin{align}
    \braket{A_\xi^2} \leq 4\mathcal{V}E,
\end{align}
which determines the scaling of the field-amplitude cutoff,
\begin{align}\label{eq:Amax}
    A_{\rm max}
    \sim O\left(
    \sqrt{\frac{8(N_c^2-1)\mathcal{V}^2E}{\varepsilon_s}}
    \right).
\end{align}

Similarly, $\braket{\Pi_\xi^2} \leq 2E$, giving the scaling of the conjugate momentum cutoff,
\begin{align}\label{eq:Pimax}
    \Pi_{\rm max}
    \sim O\left(
    \sqrt{\frac{4(N_c^2-1)\mathcal{V}E}{\varepsilon_s}}
    \right).
\end{align}

Thus to control the error of the wavefunction $\ket{\Psi}$ of order $\varepsilon_s$, the qubit cost for encoding the gauge fields ${A}_\xi$ for a given $\xi$ is
\begin{align}
    \mathcal{K} &\sim \log_2(A_{\rm max} /\delta_A)\nonumber\\
    &\sim \log_2(A_{\rm max} \Pi_{\rm max}) \sim  \log_2\left(\frac{\mathcal{V}^{3/2} E}{\varepsilon_s}\right)\,,
\end{align}
so the total number of qubits used for encoding the gauge field is
\begin{align}\label{eq:NG}
    N_G &= 2 (N_c^2-1)\mathcal{V} \mathcal{K} \nonumber\\
    &\sim 2(N_c^2-1)\mathcal{V} \log_2\left(\frac{\mathcal{V}^{3/2} E}{\varepsilon_s}\right)\,.
\end{align}

While the above estimates yield generic polynomial bounds on the discretization errors, observables satisfying the Nyquist--Shannon (NS) sampling criterion~\cite{Somma:2015bcw,Macridin:2018gdw,Macridin:2018oli,Klco:2018zqz} can converge exponentially fast in the number of digitization states $2^\mathcal{K}$ for $A_\xi$, substantially reducing the qubit cost for specific processes.

With the above digitization, the gauge field operator ${A}_\xi$ can be represented by the linear combination of Pauli strings
\begin{align}\label{eq:qA}
    {A}_\xi = - \delta_A \sum_{\mathcal{J}=0}^{\mathcal{K}-1} 2^{\mathcal{J}-1} \sigma^z_{\mathcal{J},\xi}\,.
\end{align}
The conjugate momentum operator ${\Pi}_\xi$ can be obtained by applying
the Fourier transform to the gauge-field operator ${A}_\xi$,
\begin{align}
    {\Pi}_\xi =
    \frac{\delta_\Pi}{\delta_A}
    U^\dagger_{\rm FT}(\xi) {A}_\xi U_{\rm FT}(\xi)\,,
\end{align}
where $U_{\rm FT}(\xi)$ is the approximate $\mathcal{K}$-qubit Fourier transform, with gate
complexity $O(\mathcal{K}\log_2\mathcal{K})$~\cite{Nam:2019sox}.

\vspace{0.5cm}
\noindent{\it Time evolution complexity:}
We demonstrate an efficient circuit implementation of the time-evolution
operator $\Uax(t)\equiv \exp(-\im\Hfix t)$. The residual global gauge freedom is removed by restricting the initial state to the global charge-singlet sector. Since $\Hfix$ is invariant under the residual global gauge transformation, this projection is required only during state preparation and does not introduce additional cost during subsequent Hamiltonian time evolution. The term $H_V$ contains products of the form $A\Pi A\Pi$, which are non-diagonal in both the field-amplitude and conjugate-momentum bases whenever an $A$ and a $\Pi$ operator share the same index. We therefore employ QSVT to efficiently implement these products through block encodings of the constituent operators.
\begin{figure}
    \centering
    \includegraphics[width=0.40\textwidth]{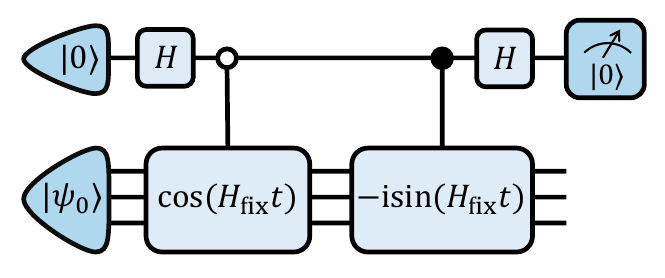}
    \caption{Circuit realizing the time evolution operator corresponding to $\Hfix$ by QSVT. The two control operations $\ket{0}\bra{0}\otimes \cos(\Hfix t)+\ket{1}\bra{1}\otimes I_{\rm sys}$ and $\ket{0}\bra{0}\otimes I_{\rm sys}-\im \ket{1}\bra{1}\otimes \sin(\Hfix t)$ are realized by the block encoding of $\Hfix$.}
    \label{fig:QSVT-circuit}
\end{figure}

QSVT implements the unitary $\Uax(t)$ through the decomposition
\begin{align}
    \Uax(t)=\cos(\Hfix t)-\im\sin(\Hfix t).
\end{align}
The resulting matrix functions act on an arbitrary initial state
$\ket{\psi_0}$ through the QSVT circuit is shown in
Fig.~\ref{fig:QSVT-circuit}. Both $\cos(\Hfix t)$ and $\sin(\Hfix t)$ admit
infinite polynomial expansions in $\Hfix$. Truncating these expansions at an appropriate degree yields finite polynomials that can be implemented efficiently using query access to the block encoding of $\Hfix$, denoted by $B_{\Hfix}$.
The terms except $H_V$ in $\Hfix$ can be block encoded by linear combination of unitaries (LCU) according to their expression, with a gate complexity $O(\mathcal{V}\mathcal{K}[\log_2(\mathcal{V}\mathcal{K})]^2)$. For $H_V$, we approximate $K^{-1}$ to error $\varepsilon_{\rm inv}$ using
QSVT, which requires
$\kappa\log(\kappa/\varepsilon_{\rm inv})$
queries to a block encoding of $K$~\cite{PRXQuantum.2.040203}, where
$\kappa=O(\mathcal{V}^{4/3})$.
Although $K^{-1}$ is generally nonlocal, $K$ itself involves only
next-to-nearest-neighbor interactions. Thus, the QSVT implementation
requires only local queries to $K$, avoiding direct long-range qubit
interactions. Each query to the block encoding of $K$ can be implemented
with gate complexity $O(\log^ 3_2\mathcal{V})$.

The query complexity of QSVT to realize $\cos(\Hfix t)$ and $\sin(\Hfix t)$ depends on the spectral norm of the Hamiltonian $\|\Hfix \|$, evolution time $t$, and the error $\varepsilon_s$ due to the cut-off of the infinite series. Specifically, QSVT constructs an approximate block encoding of $\Uax(t)$ with error $\varepsilon_s$ using a total of
\begin{align}
    \Theta\left(\|\Hfix\| t + \frac{\log(1/\varepsilon_s)}
    {\log[e+\log(1/\varepsilon_s)/(\|\Hfix\| t)]}\right)
    \label{eq:MU}
\end{align}
queries to $B_{\Hfix}$~\cite{Gilyen:2018khw,PRXQuantum.2.040203}.
Here the logarithmic dependence on $1/\varepsilon_s$ makes QSVT suitable for high-precision simulation. Assuming a moderate error tolerance, the second term in Eq.~\eqref{eq:MU} can be ignored, and the circuit depth of the QSVT-based simulation is proportional to $\|\Hfix \|$.

We thus estimate the operator norm of $\Hfix$. The leading contribution to
the upper bound on $\|\Hfix\|$ arises from $H_V$, which contains the inverse
operator $K^{-1}$. Since $K$ is positive definite on the physical subspace,
$K^{-1}$ is bounded, this gives
\begin{align}\label{eq:alphaB}
    \|H_{\rm fix}\| \leq &O\left(\frac{g^2 \mathcal{V}^{16/3} E^2}{\varepsilon_s^2}\right)\,,
\end{align}
From Eqs.~\eqref{eq:MU} and~\eqref{eq:alphaB}, together with the gate complexity of $O(\mathcal{V}^{4/3})$ to implement $B_{H_{\rm fix}}$ (up to logarithmic factors), we obtain the gate
complexity for simulating the time-evolution operator $\Uax$,
\begin{align}
    \mathcal{M}_U \sim O\left(
    \frac{t g^2 \mathcal{V}^{20/3} E^2}{\varepsilon_s^2}
    \right).
\end{align}
This complexity scales polynomially with the lattice volume $\mathcal{V}$
and the inverse simulation precision $1/\varepsilon_s$, demonstrating the
efficiency of QSVT-based simulations of dynamics in YM theory.

\vspace{0.5cm}
\noindent{\it Summary and Outlook:} 
We have established a quantum algorithmic framework for
simulating dynamics in non-Abelian Yang--Mills theories, providing a
significant step toward first-principles simulations of non-perturbative
QCD dynamics. This framework is based on fixing the maximal-tree gauge
directly in terms of the gauge field variables and performing
field-amplitude digitization. Using QSVT to implement the resulting
interaction terms, we have shown that both the qubit cost for encoding the
Yang--Mills theory and the gate complexity of its time evolution scale
polynomially with the relevant parameters. 

Looking forward, this framework provides a foundation for extracting nonperturbative dynamical quantities relevant to QCD. The gauge field variables used here are noncompact, whereas the underlying gauge transformations are generated by the compact group $SU(N_c)$. Gauge-invariant noncompact formulations have previously been developed for non-Abelian gauge theories, with Hamiltonian analyses and numerical studies providing evidence for confinement and the expected weak-coupling behavior \cite{Palumbo:1990,Diekmann:1994,Cahill:1996}. A systematic study of the continuum relation between the present gauge-fixed field-amplitude formulation and the standard compact-link formulation is therefore an important direction for future work.

Incorporating fermions into the framework is straightforward, while the treatment of chiral fermions and fermion encodings in higher dimensions remains to be explored. Future efforts can also aim to reduce the algorithmic overhead associated with the worst-case scalings of $A_{\rm max}$ and $\Pi_{\rm max}$ considered here. In particular, exploiting the faster convergence of physically relevant observables beyond the worst-case estimates could substantially reduce the required resources. Such improvements will be important for bringing real-time quantum simulations of nonperturbative QCD closer to practical implementation on fault-tolerant quantum computers. 
\vspace{0.5cm}

\noindent{\it Acknowledgments}.- We thank Christian Bauer, Andreas Kronfeld, Henry Lamm, Lingxiao Wang, Tetsuo Hatsuda and QuNu members for insightful discussions and valuable comments. This work is supported by the National Key
R\&D Program of China (Grant Nos. 2025YFA1614200), the National Science Foundation
of China under Grant Nos. 12522509, 12525508, 12475139. This research was also supported by the RIKEN TRIP initiative (RIKEN Quantum) and the UTokyo Quantum Initiative. Y.-Y. Li would like to thank the Aspen Center for Physics, which is
supported by National Science Foundation grant No. PHY-2210452, where part of this work
has been done. 

\section{END NOTE}
\noindent{\it Definition of derivatives and Gauss's law:} We define the forward and backward finite differences to approximate the derivative in NBCs
\begin{align}
    \partial^R f(n)=\frac{f(n+1)-f(n)}{a_s},
\qquad n=0,\ldots,L-2,
\end{align}
and
\begin{align}
    \partial^L h(n)=
    \begin{cases}
        \dfrac{h(0)}{a_s},&n=0,\\[2mm]
        \dfrac{h(n)-h(n-1)}{a_s},&1\le n\le L-2,\\[2mm]
        -\dfrac{h(L-2)}{a_s},&n=L-1.
    \end{cases}
\end{align}
The differences of ${A}_\mu$ and its conjugate momenta ${\Pi}^\mu$ are given by
\begin{equation}
    \begin{aligned}\label{eq:parAPI}
    &\partial_i A^a_\nu(\mathbf{x}) \to {\partial}^R_i {A}^a_\nu(n)\,\\
    &\partial_i \Pi^{\nu a}(\mathbf{x}) \to {\partial}^L_i {\Pi}^{\nu a}(n)\,,
\end{aligned}
\end{equation}
with $i=1,2,3$, and the second-order derivative $\partial_i^2$ is naturally defined by
\begin{align}
\partial_i^2 f(n)
&=
\left(\partial_i^L\partial_i^R f\right)(n)
\nonumber\\
&=
\frac{1}{a_s^2}
    \begin{cases}
    f(n+\hat{i})-f(n),
    \, n_i=0,\\[1mm]
    f(n+\hat{i})+f(n-\hat{i})-2f(n),
    \,1\leq n_i\leq L-2,\\[1mm]
    f(n-\hat{i})-f(n),
    \,n_i=L-1.
    \end{cases}
\end{align}
Hence, Gauss's law on the lattice is given by
\begin{align}
    \partial^L_i \Pi^{ib}(n) + g f^{abc} A^c_{i}(n) \Pi^{ia}(n) = 0\,.
\end{align}
Then, we separate the dependent and physical variables in Gauss's law. For the dependent variables, we replace it with $A_0$ by using secondary constraints in Tab.~\ref{tab:variables}. After doing those things, we can obtain $K A^a_0+\mathcal{Q}^a = 0$ in the main text, where $\mathcal{Q}^a(n)$ is the Gauss's law operator in terms of only physical variables, given by
($I = 1,2$)
\begin{align}\label{eq:v4-Qdef}
    &\mathcal{Q}^a = (D_I^L[A])^{ba}\Pi^{Ib}_{\rm Phys}\,.
\end{align}

\bibliography{refs} % 指向您的 refs.bib 文件

\clearpage
\onecolumngrid

\setcounter{section}{0}
\setcounter{subsection}{0}
\setcounter{equation}{0}
\setcounter{figure}{0}
\setcounter{table}{0}
\renewcommand{\thesection}{S\arabic{section}}
\renewcommand{\theequation}{S\arabic{equation}}
\renewcommand{\thefigure}{S\arabic{figure}}
\renewcommand{\thetable}{S\arabic{table}}
\renewcommand{\theHsection}{S\arabic{section}}
\renewcommand{\theHequation}{S\arabic{equation}}
\renewcommand{\theHfigure}{S\arabic{figure}}
\renewcommand{\theHtable}{S\arabic{table}}

\begin{center}
    {\large\bfseries Supplemental Material for Efficient Quantum Simulations of Yang--Mills Theory with Maximal-Tree Gauge}
\end{center}
\vspace{1em}

\def\SMINCLUDED{1}
\ifdefined\SMINCLUDED
    \let\SMfinish\relax
\else
\def\SMfinish{\bibliography{refs}\end{document}}

\documentclass[aps,pra,onecolumn,superscriptaddress,notitlepage,nofootinbib,10pt]{revtex4-1}

\usepackage{amsmath,amssymb,amsthm,amscd}
\usepackage{array}
\usepackage{color}
\usepackage{enumitem}
\usepackage{graphicx}
\usepackage{quantikz}
\usepackage{braket}
\usepackage{hyperref}
\usepackage[normalem]{ulem}
\hypersetup{colorlinks=true, linkcolor=blue, citecolor=blue, urlcolor=magenta}

\renewcommand{\thepage}{S\arabic{page}}
\renewcommand{\thesection}{S\arabic{section}}
\renewcommand{\thetable}{S\arabic{table}}
\renewcommand{\thefigure}{S\arabic{figure}}
\renewcommand{\theequation}{S\arabic{equation}}

\newtheorem{theorem}{Theorem}
\newtheorem{lemma}{Lemma}
\newtheorem{corollary}{Corollary}
\newtheorem{remark}{Remark}

\newcommand{\im}{\mathrm{i}}
\newcommand{\OO}{\mathcal{O}}
\newcommand{\V}{\mathcal V}
\newcommand{\Fp}{\mathcal F}
\newcommand{\Amax}{A_{\rm max}}
\newcommand{\Aopmax}{A_{\rm op,max}}
\newcommand{\Pimax}{\Pi_{\rm max}}
\newcommand{\Piopmax}{\Pi_{\rm op,max}}
\newcommand{\Gfp}{\Gamma_{\rm FP}}
\newcommand{\Klat}{\mathcal K_{\rm lat}}
\newcommand{\Hred}{H_{\rm red}}
\newcommand{\Hfix}{H_{\rm fix}}
\newcommand{\ad}{\operatorname{ad}}
\newcommand{\cP}{c_{\rm P}}
\newcommand{\cad}{c_{\rm ad}}
\newcommand{\poly}{\operatorname{poly}}
\newcommand{\polylog}{\operatorname{polylog}}
\newcommand{\Ran}{\operatorname{Ran}}
\newcommand{\Png}{P_{\rm ng}}
\newcommand{\ExpPsi}[1]{\left\langle #1\right\rangle_{\Psi}}
\newcommand{\Spos}{\mathcal{S}_{\rm pos}}
\newcommand{\Scol}{\mathcal{S}_{\rm col}}
\newcommand{\Slat}{\mathcal{S}_{\rm lat}}

\newcommand{\yy}[1]{\textcolor{blue}{#1}}

\begin{document}
\preprint{RIKEN-iTHEMS-Report-26}

\title{Supplemental Material for Efficient Quantum Simulations of Yang--Mills Theory with Maximal-Tree Gauge}

\author{Tianyin Li}
\email{tianyin.li@riken.jp}
\affiliation{RIKEN Center for Interdisciplinary Theoretical and Mathematical Sciences (iTHEMS), RIKEN, Wako 351-0198, Japan}

\author{Ying-Ying Li}
\email{liyingying@ihep.ac.cn}
\affiliation{Institute of High Energy Physics, Chinese Academy of Sciences, Beijing 100049, China}

\author{Xiaoyang Wang}
\email{xiaoyang.wang@riken.jp}
\affiliation{RIKEN Center for Interdisciplinary Theoretical and Mathematical Sciences (iTHEMS), RIKEN, Wako 351-0198, Japan}
\affiliation{RIKEN Center for Computational Science (R-CCS), Kobe 650-0047, Japan}

\author{Hongxi Xing}
\email{hxing@m.scnu.edu.cn}
\affiliation{State Key Laboratory of Nuclear Physics and Technology, Institute of Quantum Matter,
South China Normal University, Guangzhou 510006, China}
\affiliation{Guangdong Basic Research Center of Excellence for Structure and Fundamental Interactions of Matter,
Guangdong Provincial Key Laboratory of Nuclear Science, Guangzhou 510006, China}

\date{\today}
\maketitle
\fi

\section{Hamiltonian of a non-Abelian gauge theory in the maximal-tree gauge}\label{sec:ham}

The maximal-tree gauge condition is given by
\begin{align}
    &A^a_3(n) = 0\,,\nonumber\\
    &A^a_2(n_{12},n_3 = 0) = 0\,,\nonumber\\
    &A^a_1(n_1,n_{23} = 0) = 0\,,
\end{align}
where $n = (n_1,n_2,n_3)$ and $n_{ij} = (n_i,n_j)$.

The lattice Hamiltonian in the maximal-tree gauge is
\begin{align}
    \Hfix &= \frac{1}{2} \sum_{n} \left(\Pi^a_{I,{\rm Phys}}(n)\right)^2 + \frac{1}{2}  \sum_n \sum_{i>j} \left(F^a_{ij}(n)\right)^2 \nonumber\\
    &\quad + \left[\frac{1}{2} \sum_n \left(\Pi^a_3(n)\right)^2 + \frac{1}{2} \sum_{n_{12}} \left(\Pi^a_2(n_{12},0)\right)^2 + \frac{1}{2} \sum_{n_1}\left(\Pi^a_1(n_1,0,0)\right)^2\right]\nonumber\\
    & = H_\Pi + H_B + H_V\,.
\end{align}
Our goal is to express $\Hfix$ entirely in terms of the physical variables. The term $H_\Pi$ is already written in this form. The field strength $F^a_{ij}(n)$ appearing in $H_B$ is
\begin{align}
    &F^a_{3I} = \partial^R_3 A^a_{I,{\rm Phys}}\,,\nonumber\\
    &F^a_{IJ} = \partial^R_I A^a_{J,{\rm Phys}} - \partial^R_J A^a_{I,{\rm Phys}} + g[\ad(A_{I,{\rm Phys}})A_{J,{\rm Phys}}]^a\,,
\end{align}
where $\ad(A_{I,{\rm Phys}})$ is the adjoint representation of physical gauge field oprator $A^a_{I,{\rm Phys}}$, and $\Pi^{Ia}_{\rm Phys}(n)$ is the variable that conjugate to $A^a_{I,{\rm Phys}}$. $\partial^R_i$ and $\partial^L_i$ are forward and backward finite differences on a lattice with Neumann boundary conditions (NBCs)
\begin{align}
    \partial^R f(n)=\frac{f(n+1)-f(n)}{a_s},
\qquad n=0,\ldots,L-2,
\end{align}
and
\begin{align}
    \partial^L h(n)=
    \begin{cases}
        \dfrac{h(0)}{a_s},&n=0,\\[2mm]
        \dfrac{h(n)-h(n-1)}{a_s},&1\le n\le L-2,\\[2mm]
        -\dfrac{h(L-2)}{a_s},&n=L-1.
    \end{cases}
\end{align}
Thus, $H_B$ is also expressed entirely in terms of the physical variables.

To express $H_V$ in terms of the physical variables, we use the following four secondary constraints:
\begin{align}
    &\mathcal{G}^a(n) = \partial_i^L \Pi^{ia}(n) + g f^{abc} A_i^b(n) \Pi^{ic}(n) = 0\,,\nonumber\\
    &\Pi_1^a(n_1,n_{23} = 0) - \partial_1^R A_0^a(n_1,n_{23}=0)=0\,,\nonumber\\
    &\Pi_2^a(n_{12},n_3 = 0) - \partial_2^R A_0^a(n_{12},n_3=0) = 0\,,\nonumber\\
    &\Pi_3^a(n) - \partial^R_3 A_0^a(n) = 0 \,,
\end{align}
The term $H_V$ can then be written in terms of $A_0^a(n)$ as
\begin{align}
    H_V &= \frac{1}{2} \sum_n \left[(\partial_3^R A_0^a(n))^2 + \delta_{n_3,0} (\partial_2^R A_0^a(n))^2 + \delta_{n_2,0} \delta_{n_3,0}(\partial_1^R A_0^a(n))^2\right]\nonumber\\
    & = \frac{1}{2} \sum_n \left[-A_0^a(n)\partial^L_3\partial^R_3A_0^a(n) - \delta_{n_3,0} A_0^a(n) \partial_2^L \partial_2^R A_0^a(n) - \delta_{n_2,0}\delta_{n_3,0}A_0^a(n)\partial_1^L \partial_1^R A_0^a(n)\right]\nonumber\\
    & = \frac{1}{2} \sum_{n,m} A_0^a(n) K_{n,m} A_0^a(m)\,,
\end{align}
where we have used $\sum_n (\partial^R f(n))g(n) = -\sum_n f(n) (\partial^Lg(n))$ NBCs. With $\partial_i^2 = \partial_i^L \partial_i^R$, this gives
\begin{align}
    K = -\partial_3^2 - \delta_{n_3,0} \partial_2^2 - \delta_{n_2,0}\delta_{n_3,0} \partial_1^2\,.
\end{align}
$H_V$ depends on $A_0^a$, which is determined by Gauss's law, $\mathcal{G}^a(n) = 0$. Using the last three secondary constraints, both the $\partial^L_i \Pi^{ia}$ and $f^{abc} A^c_{i} \Pi^{ia}$ terms in Gauss's law can be rewritten in terms of the physical variables:
\begin{align}
    \partial^L_i\Pi^{ia}(n) &= -\partial_3^2 A_0^a(n) - \delta_{n_3,0} \partial_2^2A_0^a(n) - \delta_{n_2,0}\delta_{n_3,0} \partial_1^2A_0^a(n) + \partial^L_I \Pi^{Ia}_{\rm phys}\nonumber\\
    &= \sum_m K_{n,m}A_0^a(m) + \partial^L_I \Pi^{Ia}_{\rm Phys}\,,
\end{align}
and
\begin{align}
    g f^{abc}A_i^b(n) \Pi^{ic}(n) = g f^{abc} A_{I,{\rm Phys}}^b(n) \Pi^{Ic}_{\rm Phys}(n)\,.
\end{align}
Gauss's law therefore becomes
\begin{align}
    \mathcal{G}^a(n) = \sum_{m} K_{n,m} A_0^a(m) + \mathcal{Q}^a(n) = 0\,,
\end{align}
where
\begin{align}
    \mathcal{Q}^a(n) = \partial^L_I \Pi^{Ia}_{\rm Phys} + g f^{abc} A_{I,{\rm Phys}}^b(n) \Pi^{Ic}_{\rm Phys}(n)\,.
\end{align}
Solving Gauss's law for $A_0^a$ gives
\begin{align}
    A_0^a(n) = -\sum_m(K^{-1})_{n,m} \mathcal{Q}^a(m)\,.
\end{align}
Substituting $A_0^a(n)$ into $H_V$, we obtain
\begin{align}
    H_V = \frac{1}{2} \sum_{n,m} [(K^{-1})_{n,m} \mathcal{Q}^a(m)] \mathcal{Q}^a(n) = \frac{1}{2} \sum_{n,m}(\mathcal{Q}^a(n))^\dagger (K^{-1})_{n,m} \mathcal{Q}^a(m)\,,
\end{align}
where we have used $(\mathcal{Q}^{a}(n))^\dagger = \mathcal{Q}^a(n)$.

The gauge field needs to be digitized because the Hilbert space of quantum computers is finite. In the main text, we truncate the eigenbasis of the gauge field $A_\xi$, where index $\xi$ denotes all indices of the gauge fields, including the color index, the index labeling the two physical operators, and the spatial-coordinate index specifying the lattice site. After digitization, we have
\begin{align}
    {A}_\xi \ket{A}_\xi = A\ket{A}_\xi\,,
\end{align}
where $-A_{\rm max} \leq A = -A_{\rm max}+\lambda \delta_A \leq A_{\rm max}$, with $\lambda = 0,1,2,...,2^\mathcal{K}-1$, and $\delta_A = 2A_{\rm max}/(2^\mathcal{K}-1)$ is the spacing between adjacent eigenvalues of ${A}_\xi$. On the other hand, the conjugate-momentum basis of ${\Pi}_\xi$, denoted by
$\ket{\Pi}_\xi$, is obtained from the $\ket{A}_\xi$ basis by a discrete
quantum Fourier transformation. The truncation scale of ${\Pi}_\xi$, denoted by $\Pi_{\rm max}$, is of $O(1/\delta_A)$, with a spacing $\delta_\Pi \sim 1/A_{\rm max}$.

We characterize the truncation error $\varepsilon_s$ of the wavefunction $\ket{\Psi}$ to estimate the required number of qubits. Generally, a wavefunction $\ket{\Psi}$ satisfies
\begin{align}
    \sum_A|\langle \Phi_A| \Psi \rangle|^2 = \begin{cases}
        1-\varepsilon_s \,, \ {\rm if}\ |A_{\xi = 1}|,...,|A_{\xi = \xi_{\rm max}}| \leq A_{\max}\,,\\
        \varepsilon_s\,, \ {\rm others}\,,
    \end{cases}
\end{align}
where $\xi_{\rm max}$ is total number of collective indexes, and $\ket{\Phi_A} \equiv \otimes_{\xi} \ket{A}_{\xi}$. The basis $\ket{\Phi_A}$ spans the Hilbert space of gauge fields, which is denoted by $\mathcal{H} = {\rm span}\{\ket{\Phi_A}\}_{\forall A}$. There is also a similar truncation error of the wavefunction in the $\ket{\Phi_{\Pi}} = \otimes_\xi \ket{\Pi}_\xi$ basis. From the main text, we know that the truncation scale $A_{\rm max}$ and $\Pi_{\rm max}$ are determined by the desired $\varepsilon_s$, number of lattice sites $\mathcal{V}$, and energy scale $E$, 
which is given by
\begin{align}\label{eq:SM-APim}
    &A_{\rm max}
    \sim O\left(
    \sqrt{\frac{8(N_c^2-1)\mathcal{V}^2E}{\varepsilon_s}}
    \right), \nonumber\\
    &\Pi_{\rm max}
    \sim O\left(
    \sqrt{\frac{4(N_c^2-1)\mathcal{V}E}{\varepsilon_s}}
    \right).
\end{align}

\section{Condition number $\kappa$ of $K$}

The key step in representing $A_0^a(n)$ on a quantum computer is to implement $K^{-1}$ as a quantum circuit. The corresponding complexity depends on the condition number of $K$, which we write as $\kappa = \alpha_K \Gamma$. Here, $\alpha_K$ is an upper bound on the largest eigenvalue of $K$, while $\Gamma^{-1}$ is a lower bound on its smallest eigenvalue. In this section, we derive the scaling of $\kappa$.

To this end, we regard $K$ as a matrix by choosing the basis $\ket{n}_{\Spos}$ for the position space $\Spos$. Hereafter, we abbreviate $\ket{n}_{\Spos}$ as $\ket{n}$ and write $\ket{n} = \ket{n_1} \ket{n_2} \ket{n_3}$. An arbitrary function $\omega(n)$ on the lattice can then be represented by the state
\begin{align}
    \ket{\omega} = \sum_n \omega(n) \ket{n}\,.
\end{align}
The corresponding inner product is
\begin{align}
    \braket{\omega_1|\omega_2} = \sum_n \omega_1^*(n) \omega_2(n)\,.
\end{align}
The action of $\partial^R_i$ on $\ket{\omega}$ is
\begin{align}\label{eq:paromega}
    \partial^R_i \ket{\omega} = \sum_{n=0}^{L-2} \partial^R_i \omega(n) \ket{n} = \sum_{n=0}^{L-2} \frac{1}{a_s}[\omega(n+1)-\omega(n)]\ket{n}\,.
\end{align}
Thus, $\bra{n}\partial^R_i \ket{\omega} = \partial^R_i \omega(n)$. Equation~\eqref{eq:paromega} then determines the action of $\partial^R_i$ on the basis state $\ket{n}$:
\begin{align}\label{eq:p3op}
    \partial_i^R \ket{n} =\begin{cases}
        -\frac{1}{a_s}\ket{0},\ {\rm if} \ n=0\,,\\
        \frac{1}{a_s} (\ket{n-\hat{i}}-\ket{n}), \ {\rm if}\ 1\leq n_i \leq L-2,\\
        \frac{1}{a_s}\ket{n-\hat{i}},\ {\rm if} \ n=L-1.
    \end{cases} 
\end{align}
Without loss of generality, we set $a_s=1$ from now on. The adjoint $(\partial^R_i)^\dagger$ has the simpler form
\begin{align}
    (\partial^R_i)^\dagger\ket{n} = -\partial^L_i\ket{n} = \ket{n+\hat{i}} - \ket{n},\ 0\leq n_i \leq L-2\,.
\end{align}
Thus, $\partial^R_i$ is a matrix with at most two nonzero elements in each row. With these conventions, the quadratic form $\braket{\omega|K|\omega}$ is
\begin{align}\label{eq:inK}
    \braket{\omega |K|\omega} &= \braket{\omega|(\partial_3^R)^\dagger\partial_3^R|\omega} + \braket{\omega|\delta_{n_3,0}(\partial_2^R)^\dagger\partial_2^R|\omega} + \braket{\omega|\delta_{n_{2},0} \delta_{n_3,0}(\partial_1^R)^\dagger\partial_1^R|\omega}
    \nonumber\\
    &=\sum_n |\partial^R_3 \omega(n)|^2 + \sum_{n_{12}} |\partial_2^R \omega(n_{12},0)|^2 + \sum_{n_1} |\partial_1^R \omega(n_{1},0,0)|^2\,.
\end{align}

Physical states satisfy $\sum_n \mathcal{Q}(n) \ket{\rm Phys} = 0$. Because $K$ acts on $\mathcal{Q}$ in $H_V$, it is sufficient to consider functions $\omega(n)$ satisfying $\sum_{n}\omega(n) = 0$. For any such function,
\begin{align}
    \braket{\omega|\omega} = \frac{1}{2\mathcal{V}}\sum_{n,m}|\omega(n)-\omega(m)|^2\,.
\end{align}
Let $n\equiv l_1 \to l_2\to,...,\to l_P\equiv m$ be a shortest path along the maximal tree, with each step connecting neighboring lattice sites, and define $\Omega_p = \omega(l_{p+1}) - \omega(l_p)$. Then
\begin{align}
    \omega(n) - \omega(m) = \sum_{p=1}^{P-1} \Omega_p\,.
\end{align}
The Cauchy--Schwarz inequality gives
\begin{align}
    |\omega(n)-\omega(m)|^2 \leq (P-1) \sum_{p=1}^{P-1} |\Omega_p|^2\,.
\end{align}
For any pair of sites $n$ and $m$ on a lattice of volume $\mathcal{V} = L^3$, the path length is at most $5L-3$, so $P-1 \leq 5L-3$. Moreover,
\begin{align}
    \sum_{p=1}^{P-1} |\Omega_p|^2 = \sum_{p=1}^{P-1} |\omega(l_{p+1})-\omega(l_p)|^2 \leq \braket{\omega|K|\omega}\,,
\end{align}
because $\braket{\omega|K|\omega}$ contains the contributions from all pairs of adjacent lattice sites along the maximal tree. Therefore,
\begin{align}
    |\omega(n)-\omega(m)|^2 \leq (5L-3) \braket{\omega|K|\omega}\,.
\end{align}
It follows that
\begin{align}
    \braket{\omega|\omega} &= \frac{1}{2\mathcal{V}} \sum_{n,m} |\omega(n)-\omega(m)|^2 \nonumber\\
    &\leq \frac{1}{2\mathcal{V}} \sum_{n,m} (5L-3) \braket{\omega|K|\omega}\nonumber\\
    &= \frac{(5L-3)\mathcal{V}}{2} \braket{\omega|K|\omega}\,.
\end{align}
We therefore obtain
\begin{align}
    \Gamma = \frac{(5L-3)\mathcal{V}}{2} \sim O(\mathcal{V}^{4/3})\,.
\end{align}

To obtain the upper bound, we apply
\(|u-v|^2\le2|u|^2+2|v|^2\) in every term of
Eq.~\eqref{eq:inK}.  Each site occurs in at most four displayed
differences.  Hence
\begin{align}
    \langle\omega|K|\omega\rangle
    \le8\|\omega\|^2 < 12\|\omega\|^2.
\end{align}
We use $\alpha_K = 12$ to consist with $\alpha_K$ in \ref{subsubsec:block_KA}. $\alpha_K = 12 \sim O(1)$, and the condition number scales as
\begin{align}
    \kappa = \alpha_K \Gamma \sim O(\mathcal{V}^{4/3})\,.
\end{align}

\section{Block encoding of $H_{\rm fix}$}

In this section, we suppress the label ``Phys,'' because only the physical variables are encoded on the quantum computer; for example, $\Pi^a_{I,\rm{Phys}} \equiv \Pi^a_I$. The Hamiltonian was given in the main text and in Sec.~\ref{sec:ham}. For convenience, we reproduce it here:
\begin{align}
    \Hfix &= H_\Pi + H_B + H_V\nonumber\\
    &= \frac{1}{2}\sum_{n} ({\Pi}^a_{I}(n))^2 + \frac{1}{2} \sum_{n} \sum_{i>j} (F^a_{ij}(n))^2 + \frac{1}{2} \sum_{n,m} [\mathcal{Q}^a(n)]^\dagger (K^{-1})_{n,m} \mathcal{Q}^a(m)\,,
\end{align}
where $I,J = 1,2$, with
\begin{align}\label{eq:defFIJ}
    &F^a_{3I} = \partial^R_3 A^a_I\,,\nonumber\\
    &F^a_{IJ} = \partial^R_I A^a_J - \partial^R_J A^a_I + g[\ad(A_I)A_J]^a\,.
\end{align}
and
\begin{align}
    &K
    =
    -\partial_3^2
    -
    \delta_{n_3,0}\partial_2^2
    -
    \delta_{n_2,0} \delta_{n_3,0} \partial_1^2\,,\nonumber\\
    &\mathcal{Q}^a = (D_I^L[A])^{ba}\Pi^{Ib} = \partial^L_I \Pi^{Ia} + g f^{bac} A^c_{I} \Pi^{Ib} \equiv \partial^L_I \Pi^{Ia} + g[\ad(A_I)\Pi^I]^a\,.
\end{align}

The Hamiltonian is built from the summation of local operators. A convenient way to deal with the summation of $n$ and $a$ in the Hamiltonian is to define an extended linear map $\widetilde{A}_I$ that coherently encodes the local operators $A_I^a(n)$:
\begin{align}\label{eq:tildeA}
    \tilde{A}_I \ket{\psi} = \sum_{n,a} \ket{n} \ket{a}_{\Scol} A_{I}^a(n) \ket{\psi}\,.
\end{align}
Here, $\ket{\psi}$ is a state in Hilbert space $\mathcal{H}$ ($\mathcal{H}$ is spanned by $\ket{\Phi_A}$), and $\ket{a}_{\Scol}$ is a basis state of the color space $\Scol$. From now on, we suppress the lower index $\Scol$ and write $\ket{a}_{\Scol}$ as $\ket{a}$. As an example, we show how we can obtain $\sum_{n} (\partial^R_I A_J^a)^2$ term in the Hamiltonian by using $\tilde{A}_I$. Firstly, we act $\partial^R_I$ on $\tilde{A}_J$
\begin{align}\label{eq:pat}
    \partial^R_I \tilde{A}_J \ket{\psi} &= \partial^R_I \sum_{n,a} \ket{n} \ket{a} A^a_J(n) \ket{\psi}\nonumber\\
    &= \sum_{n,a} \ket{n} \ket{a} \partial^R_IA^a_J(n) \ket{\psi}\,,
\end{align}
where the second line of the above equation can be obtained by using Eq~\eqref{eq:p3op}. We then consider the conjugation of Eq.~\eqref{eq:pat}
\begin{align}\label{eq:patd}
    \left(\partial^R_I \tilde{A}_J \ket{\psi}\right)^\dagger = \left[\sum_{n,a} \ket{n} \ket{a} \partial^R_IA^a_J(n)\right]^\dagger = \sum_{n,a} \bra{n} \bra{a} \bra{\psi} \partial_I^R A_J^a(n)\,.
\end{align}
Thus, the summation of $a$ and $n$ is automatic if we consider the inner product between Eq.~\eqref{eq:pat} and Eq.~\eqref{eq:patd}
\begin{align}
    \left(\partial^R_I \tilde{A}_J \ket{\psi}\right)^\dagger \partial^R_I \tilde{A}_J \ket{\psi} = \sum_n \bra{\psi} (\partial_I \tilde{A}_J^a(n))^2 \ket{\psi}
\end{align}
For the same reason, we need to consider $\ad(A_I)$ as an operator on $\Spos\otimes\Scol\otimes \mathcal{H}$ space. In the computational basis $\ket{n} \ket{a} \ket{\Phi_A}$, $\ad(A_I)$ is given by
\begin{align}\label{eq:adAI}
    \ad(A_I)
    =
    \sum_{n,a,b,c}
    f^{abc}\ket{n}\ket{a} \bra{n}\bra{c}
    \otimes A_I^b(n).
\end{align}

The Hamiltonian is a function of elementary operators $A^a_I$, $\Pi^{Ia}$, and $\partial^R_i$. Once elementary operators have been block encoded, block encodings of composite operators such as $F_{ij}^a$, $K$, and $\mathcal{Q}^a$ can be constructed by multiplying and linearly combining the elementary block encodings. Finally, we obtain a block encoding of $\Hfix$ by linearly combining $H_\Pi$, $H_B$, and $H_V$ via a linear combination of unitaries (LCU).

We now construct a block encoding of $\Hfix$ in this unified basis. In Sec.~\ref{sec:block_elemetry}, we first block encode the elementary operators $\partial^R_i$, $A^a_I$, and $\Pi_a^I$. In Sec.~\ref{sec:block_KQ}, we then construct block encodings of composite operators such as $K$, $K^{-1}$, and $\mathcal{Q}$. Finally, in Sec.~\ref{sec:blHfix}, we construct the block encoding of $\Hfix$ and estimate its operator norm $\|\Hfix\|$.

\subsection{Block encoding of elementary operators}\label{sec:block_elemetry}

A unitary \(U_X\) is an \((\alpha_X,a_X,\epsilon_X)\)-block encoding of \(X\) if
\begin{align}
    \left\|
    X-\alpha_X
    (\bra{0^{a_X}}\otimes I)U_X
    (\ket{0^{a_X}}\otimes I)
    \right\|
    \le\epsilon_X\,,
    \label{eq:v4-BE-definition}
\end{align}
where $a_X$ is the number of qubits in the ancilla register.

\subsubsection{Block encoding of \texorpdfstring{\(\partial^R_i\)}{partial}}\label{subsubsec:block_partial}

We first construct the block encoding \(U_{\partial_i}\) of
\(\partial_i^R\). The forward-difference operator acts as the identity on the color and Hilbert spaces:
\begin{align}
    \partial_i^R
    =
    \partial_{i,{\Spos}}^R\otimes I_{\Scol}\otimes I_{\mathcal{H}}.
    \label{eq:v4-partial-pos-color}
\end{align}
According to Eq.~\eqref{eq:p3op},
\begin{align}
    &(\partial_{3,\Spos}^R)^\dagger
    =
    (T_3-I_{\Spos}) P_3\,,
    \nonumber\\
    &T_3\ket{n_1,n_2,n_3}
    =
    \ket{n_1,n_2,(n_3+1)\bmod L}\,,\nonumber\\
    &P_3= I_1 \otimes I_2 \otimes (I_3 - \ket{L-1}\bra{L-1}_3)\,.
    \label{eq:v4-shift-diff}
\end{align}
Here, we use $\partial^R_3$ as an example; $\partial^R_1$ and $\partial^R_2$ can be represented in the same way. Both the shift operator $T_3$ and the identity are unitary. The projector $P_3$ can be expressed as a linear combination of the identity and the reflection operator $R_3$:
\begin{align}\label{eq:PR3}
    &P_3 = \frac{1}{2} (I_3 + R_3)\,,\nonumber\\
    &R_3 = I_3 - 2\ket{L-1} \bra{L-1}_3 = I_3 - 2\ket{11,...1}\bra{11,...,1}_3\,,
\end{align}
where $R_3$ is unitary. We assume $L = 2^m$ and identify $\ket{0}_3$ with the $m$-qubit state $\ket{0^m}_3$. Thus, $\partial_3^R$ can be decomposed in terms of the three unitary operators $T_3$, $I_3$, and $R_3$.

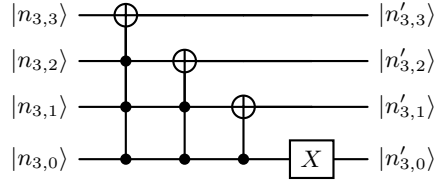
\begin{figure}[tb]
    \centering
    \begin{quantikz}[row sep=0.28cm,column sep=0.45cm]
        \lstick{$\ket{n_{3,3}}$}
        & \targ{}
        & \qw
        & \qw
        & \qw
        & \rstick{$\ket{n'_{3,3}}$} \\
        \lstick{$\ket{n_{3,2}}$}
        & \ctrl{-1}
        & \targ{}
        & \qw
        & \qw
        & \rstick{$\ket{n'_{3,2}}$} \\
        \lstick{$\ket{n_{3,1}}$}
        & \ctrl{-2}
        & \ctrl{-1}
        & \targ{}
        & \qw
        & \rstick{$\ket{n'_{3,1}}$} \\
        \lstick{$\ket{n_{3,0}}$}
        & \ctrl{-3}
        & \ctrl{-2}
        & \ctrl{-1}
        & \gate{X}
        & \rstick{$\ket{n'_{3,0}}$}
    \end{quantikz}
    \caption{Conceptual circuit for
    $T_3:\ket{n_3}\mapsto
    \ket{n_3+1\pmod{16}}$ with four qubits. The first column flips the top qubit
    only when all three lower bits are one; subsequent columns propagate the
    carry downward, and the final $X$ flips the last qubit.  The
    same pattern extends to $m$ qubits.}
    \label{fig:v4-modular-increment}
\end{figure}

We first construct the quantum circuit for $T_3$. Write
\begin{align}
    \ket{n_3}
    =
    \ket{n_{3,m-1}\cdots n_{3,1}n_{3,0}},
    \qquad
    m=\log_2L.
\end{align}
Then \(T_3\) is exactly the reversible modular increment
\begin{align}
    T_3 :
    \ket{n_3}
    \longmapsto
    \ket{n_3+1\pmod{2^{m}}}.
    \label{eq:v4-increment-action}
\end{align}
Figure~\ref{fig:v4-modular-increment} explicitly illustrates the construction of $T_3$ for four position qubits, with computational-basis labels $0,1,...,15$. A standard exact decomposition of a \(j\)-controlled \(X\) gate without ancilla qubits uses \(O(j^2)\) one- and two-qubit gates. Summing this conservative bound over \(j=1,\ldots,m-1\) gives the elementary-gate complexity of $T_3$:

\begin{align}
    \mathcal{M}(T_i)
    \sim O(m^3)
    =O(\log_2^3 L) \sim O(\log_2^3 \mathcal{V}).
    \label{eq:v4-increment-resources}
\end{align}
The gate count is therefore polylogarithmic in $\mathcal{V}$.

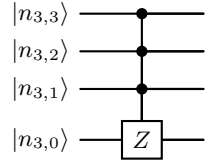
\begin{figure}[tb]
    \centering
    \begin{quantikz}[row sep=0.35cm, column sep=0.55cm]
        \lstick{$\ket{n_{3,3}}$} & \ctrl{3} & \qw \\
        \lstick{$\ket{n_{3,2}}$} & \ctrl{2} & \qw \\
        \lstick{$\ket{n_{3,1}}$} & \ctrl{1} & \qw \\
        \lstick{$\ket{n_{3,0}}$} & \gate{Z} & \qw
    \end{quantikz}
    \caption{Circuit implementing the reflection operator
    $R_3=I_3-2\ket{1111}\!\bra{1111}$ for $m=4$ position
    qubits, corresponding to $L=16$. The four-qubit
    multi-controlled-$Z$ gate changes the phase only of the
    state $\ket{1111}$.}
    \label{fig:MCZ-R3}
\end{figure}

We next construct the circuit for $R_3$. The reflection operator $I_3 - 2\ket{1}^{\otimes m}$ selectively reverses the phase of the $m$-qubit all-ones state $\ket{1}^{\otimes m}$. At the circuit level, it is implemented by an $m$-qubit multi-controlled $Z$ (MCZ) gate, which is shown in Fig.~\ref{fig:MCZ-R3}. Its gate complexity is $O(\log^2_2 \mathcal{V})$.

Using LCU to combine $I_3$ and $R_3$, we obtain a $(1,1,0)$ block encoding $U_{P_3}$ of $P_3$. Similarly, LCU gives a $(2,1,0)$ block encoding $U_{T_3}$ of $(T_3-I_{\Spos})$. The resulting $(\alpha_\partial,a_\partial,0)$ block encoding $U_{\partial_3}$ of $\partial^R_3$ is
\begin{align}
    U_{\partial_3} = U_{P_3}^\dagger U_{T_3}^\dagger \,,
\end{align}
with
\begin{align}
    &\alpha_\partial=2\,,\nonumber\\
    &a_\partial = 2\,.
\end{align}
$U_{\partial_1}$ and $U_{\partial_2}$ can be constructed analogously. The gate complexity of $U_{\partial_i}$ is determined by that of $U(T_i)$; hence,
\begin{align}
    \mathcal{M}(U_{\partial_i}) \sim \mathcal{M}(T_i) \sim O(\log_2^3\mathcal{V})\,.
\end{align}

Moreover, \(U_{\partial_i}^\dagger\) block encodes \((\partial_i^R)^\dagger=-\partial_i^L\) with the same normalization; a global minus sign therefore gives a block encoding of \(\partial_i^L\) without another construction.

\subsubsection{Block encoding of gauge potential \texorpdfstring{$A^a_I$}{A}}

Another elementary operator is the gauge potential $A^a_I(n)$. We first construct a block encoding of $A^a_I(n)$ and then one of $\tilde{A}_I = \sum_{n,a} \ket{n}\ket{a} A^a_I(n)$, which was discussed in Eq.~\eqref{eq:tildeA}. We also construct a block encoding of $\bar{A}^a_I = \sum_{n}\ket{n}\bra{n} A^a_I(n)$, which is a crucial step in constructing the block encoding of $\ad(A_I)$, as shown in Eq.~\eqref{eq:v4-adA-Mb-decomp}.

We begin with the following representation of $A^a_I(n)$:
\begin{align}
    A_\xi
    =
    -\delta_A
    \sum_{\mathcal J=0}^{\mathcal{K}-1}
    2^{\mathcal J-1}
    Z_{\mathcal J,\xi},
    \qquad
    Z_{\mathcal J,\xi}
    \equiv\sigma^z_{\mathcal J,\xi}\,,
    \label{eq:v4-Aibn-Pauli}
\end{align}
where $A_{\rm max}$ is the cutoff of the eigenvalue of the gauge field operator $A_\xi$, and $\delta_A = 2A_{\rm max}/(2^\mathcal{K}-1)$ because we use $\mathcal{K}$ qubits to encode every $A_\xi$ for a given $\xi$. Here, the $Z$ operator acts on the qubit specified by the indices $\mathcal{J}$ and $\xi = (I,a,n)$.

Define the positive LCU weights and their one-norm by
\begin{align}
    c_{\mathcal J}
    :=\delta_A2^{\mathcal J-1},
    \qquad
    \sum_{\mathcal J=0}^{\mathcal{K}-1}c_{\mathcal J}
    =\delta_A\left(2^{\mathcal{K}-1}-\frac12\right) = A_{\rm max}\,.
    \label{eq:v4-alpha-A}
\end{align}
To complete the block encoding of $A_\xi$, we construct a preparation gate ${\rm PREP}_A$ and a selection gate ${\rm SELECT}_A$. The action of ${\rm PREP}_A$ on the ancilla state $\ket{0^{a_A}}$ is
\begin{align}
    \operatorname{PREP}_{A}\ket{0^{a_A}}
    &=
    \sum_{\mathcal J=0}^{\mathcal{K}-1}
    \sqrt{\frac{c_{\mathcal J}}{A_{\rm max}}}
    \ket{{\mathcal J}},\nonumber\\
    &= \frac{1}{\sqrt{2^{\mathcal{K}}-1}}
    \sum_{\mathcal J=0}^{\mathcal{K}-1}
    2^{\mathcal J/2}\ket{{\mathcal J}}\,,
    \label{eq:v4-PREP-A-geometric}
\end{align}
where the selector register contains 
\begin{align}
    a_A = \log_2\mathcal{K}
\end{align}
qubits. Regarding $\ket{\mathcal{J}_{a_A-1},...,\mathcal{J}_0}$ as a binary string and writing
\(\mathcal J = \sum_{r=0}^{a_A-1}2^r {\mathcal J}_r\), we obtain
\begin{align}
    \operatorname{PREP}_{A}
    &=
    \bigotimes_{r=0}^{a_A-1}R_y(2\theta_r),
    &
    \tan\theta_r&=2^{2^{r-1}},
    \label{eq:v4-PREP-A-factorized}
    \\
    R_y(2\theta_r)\ket0
    &=
    \frac{\ket0+2^{2^{r-1}}\ket1}
    {\sqrt{1+2^{2^r}}}.
    \nonumber
\end{align}
Thus \(\operatorname{PREP}_A\) needs exactly \(a_A\) single-qubit rotations.  Figure~\ref{fig:v4-PREP-A-small} shows the smallest
nontrivial example, \(\mathcal{K}=4\).
\begin{figure}[tb]
    \centering
    \begin{quantikz}[row sep=0.38cm,column sep=0.65cm]
        \lstick{$\ket{\mathcal{J}_1}$: $\ket0$}
        & \gate{R_y(2\arctan 2)}
        & \rstick{$\displaystyle
        \frac{\ket0+2\ket1}{\sqrt5}$} \\
        \lstick{$\ket{\mathcal{J}_0}$: $\ket0$}
        & \gate{R_y(2\arctan\sqrt2)}
        & \rstick{$\displaystyle
        \frac{\ket0+\sqrt2\ket1}{\sqrt3}$}
    \end{quantikz}
    \caption{Exact \(\operatorname{PREP}_A\) circuit for \(\mathcal{K}=4\).  The
    output is
    \((\ket{00}+\sqrt2\ket{01}+2\ket{10}+2\sqrt2\ket{11})/\sqrt{15}\),
    as required by Eq.~\eqref{eq:v4-PREP-A-geometric}.  For
    \(\mathcal{K}=2^{a_A}\), one adds one independent rotation for every qubit according to Eq.~\eqref{eq:v4-PREP-A-factorized}.}
    \label{fig:v4-PREP-A-small}
\end{figure}
The SELECT oracle is
\begin{align}
    \operatorname{SELECT}_{A_\xi}
    =
    \sum_{\mathcal J}
     \ket{\mathcal{J}}\bra{\mathcal{J}} \otimes
    \left(-Z_{\mathcal J,
    \xi}\right).
    \label{eq:v4-SELECT-Aib}
\end{align}
The SELECT operation can be implemented directly using controlled Pauli
gates. Figure~\ref{fig:v4-SELECT-A-small} shows the circuit for ${\rm SELECT}_{A_\xi}$ at fixed $\xi$ and $\mathcal{K} = 4$. In general, the construction requires $O(\mathcal{K})$ $Z$ gates, each controlled by $O(\log_2 \mathcal{K})$ qubits. The total elementary-gate count for ${\rm SELECT}$ is therefore $O(\mathcal{K} \log^2 \mathcal{K})$.
\begin{figure}[tb]
    \centering
    \begin{quantikz}[row sep=0.25cm,column sep=0.42cm]
        \lstick{$\ket{{\mathcal J}_0}$}
        & \octrl{2} & \ctrl{3}  & \octrl{4} & \ctrl{5} & \qw \\
        \lstick{$\ket{{\mathcal J}_1}$}
        & \octrl{1} & \octrl{2} & \ctrl{3}  & \ctrl{4} & \qw \\
        \lstick{qubit ($\xi$, $\mathcal{J}=0$)}
        & \gate{-Z} & \qw       & \qw       & \qw      & \qw \\
        \lstick{qubit ($\xi$, $\mathcal{J}=1$)}
        & \qw       & \gate{-Z} & \qw       & \qw      & \qw \\
        \lstick{qubit ($\xi$, $\mathcal{J}=2$)}
        & \qw       & \qw       & \gate{-Z} & \qw      & \qw \\
        \lstick{qubit ($\xi$, $\mathcal{J}=3$)}
        & \qw       & \qw       & \qw       & \gate{-Z}& \qw
    \end{quantikz}
    \caption{Direct controlled-Z implementation of
    $\operatorname{SELECT}_{A_\xi}$ for given $\xi$ and
    $\mathcal{K}=4$.}
    \label{fig:v4-SELECT-A-small}
\end{figure}

The gates ${\rm PREP}_{A}$ and ${\rm SELECT}_{A_\xi}$ yield the $(\alpha_A = A_{\rm max},a_A = \log_2\mathcal{K},0)$ block encoding of $A_\xi$,
\begin{align}
    U_{A_\xi}
    =
    \operatorname{PREP}_{A}^\dagger
    \operatorname{SELECT}_{A_\xi}
    \operatorname{PREP}_{A}
    \label{eq:v4-UAib}
\end{align}
which satisfies
\begin{align}
    \bra{0^{a_A}}U_{A_\xi}
    \ket{0^{a_A}}
    =
    \frac{1}{A_{\rm max}}
    A_\xi\,.
    \label{eq:v4-UAib-check}
\end{align}
The complexity of $U_A$ is determined by that of ${\rm SELECT}_{A_\xi}$, giving
\begin{align}
    \mathcal{M}(U_{A_\xi}) \sim O(\mathcal{K} \log_2^2\mathcal{K})\,.
\end{align}

Next, a block encoding of $\tilde{A}_I = \sum_{n,a} \ket{n}\ket{a} A_\xi$ can be constructed using ${\rm SELECT}_{\tilde{A}_I}$ and ${\rm PREP}_{\tilde{A}}$, where
\begin{align}
    {\rm SELECT}_{\tilde{A}_I} = \sum_{n,a}\ket{n}\ket{a} \bra{n}\bra{a} U_{A_\xi }\,,
\end{align}
and
\begin{align}
    {\rm PREP}_{\tilde{A}} \ket{0} = \sum_{n}\sum_a \frac{1}{\sqrt{\mathcal{V}N_{\rm adj}}}\ket{n}\ket{a}\,.
\end{align}
Their action is
\begin{align}
    U_{\tilde{A}_I} \ket{0} = {\rm SELECT}_{\tilde{A}_I} {\rm PREP}_{\tilde{A}}\ket{0}=\sum_{n}\sum_a \frac{1}{\sqrt{\mathcal{V}N_{\rm adj}}}\ket{n}\ket{a} U_{A_\xi}\,.
\end{align}
Thus, $U_{\tilde{A}_I}$ is a $(\alpha_{\tilde{A}},a_{\tilde{A}},0)$ block encoding of $\tilde{A}_I$, with
\begin{align}
    &\alpha_{\tilde{A}} = \sqrt{\mathcal{V}N_{\rm adj}} A_{\rm max}\sim O(\mathcal{V}^{1/2} A_{\rm max})\,,\nonumber\\
    &a_{\tilde{A}} \sim \log_2{\mathcal{K}}\,.
\end{align}
The gate complexity of $U_{\tilde{A}_I}$ has the same scaling as that of $U_{\bar{A^a_I}}$; hence,
\begin{align}
    \mathcal{M}(U_{\tilde{A}_I}) \sim O(\mathcal{V}\mathcal{K} \log_2^2(\mathcal{V}\mathcal{K}))\,.
\end{align}

Finally, we construct a block encoding of $\bar{A}^a_I = \sum_{n}\ket{n}\bra{n} A_\xi$ by controlling $U_{A_\xi}$ with the $\ket{n}$ register:
\begin{align}
    U_{\bar{A}^a_I} = \sum_n \ket{n}\bra{n} U_{A_\xi}\,.
\end{align}
One can verify that $U_{\bar{A}^a_I}$ is a $(\alpha_A,a_A,0)$ block encoding of $\bar{A}^a_I$. Its implementation cost on a quantum computer is
\begin{align}
    \mathcal{M}(U_{\bar{A}^a_I}) \sim O\left(\mathcal{V}\mathcal{K}\log_2^2 (\mathcal{V}\mathcal{K})\right)\,.
\end{align}

\subsubsection{Block encoding of the independent conjugate momentum}

We next block encode the physical conjugate momentum $\Pi_\xi \equiv \Pi^{Ia}(n)$, denoting its block encoding by $U_\xi$. It is obtained by Fourier transforming $U_{A_\xi}$:
\begin{align}
    U_{\Pi_\xi} = U^{\dagger}_{\rm FT}(\xi) U_{A_\xi} U_{\rm FT}(\xi)\,,
\end{align}
where $U_{\rm FT}(\xi)$ is the Fourier-transform circuit acting on the qubits labeled by $\xi$, and we have $\Pi_{\rm max}/\delta_\Pi = A_{\rm max}/\delta_A$ here. Its complexity is $O(\mathcal{K}\log_2\mathcal{K})$. One can verify that $U_{\Pi_\xi}$ is an $(\alpha_\Pi = \Pi_{\rm max},a_A,0)$ block encoding of $\Pi_\xi$.

We also require a block encoding of $\tilde{\Pi}^I = \sum_{n,a} \ket{n}\ket{a} \Pi^{Ia}(n)$. It can be constructed by replacing ${\rm SELECT}_{\tilde{A}_I}$ in $U_{\tilde{A}_I}$ with ${\rm SELECT}_{\tilde{\Pi}^I}$, where
\begin{align}
    {\rm SELECT}_{\tilde{\Pi}^I} &= \sum_{n,a}\ket{n}\ket{a} \bra{n}\bra{a} U_{\Pi_\xi}\nonumber\\
    &=\sum_{n,a}\ket{n}\ket{a} \bra{n}\bra{a} U^\dagger_{\rm FT}(\xi) U_{A_\xi} U_{\rm FT}(\xi)\,.
\end{align}
Thus, $U_{\tilde{\Pi}^I} = {\rm SELECT}_{\tilde{\Pi}^I}{\rm PREP}_{\tilde{A}}$ is an $(\alpha_{\tilde{\Pi}},a_{\tilde{\Pi}},0)$ block encoding of $\tilde{\Pi}^I$, with
\begin{align}
    &\alpha_{\tilde{\Pi}} = \sqrt{\mathcal{V}N_{\rm adj}} \Pi_{\rm max}\sim O(\mathcal{V}^{1/2}\Pi_{\rm max})\,,\nonumber\\
    &a_{\tilde{\Pi}} \sim O( \log_2\mathcal{K})\,.
\end{align}

Unlike $U_{\tilde{A}_{I}}$, the $\mathcal{K}$-qubit Fourier-transform circuit $U_{\rm FT}$ must also be controlled by the register $\ket{n}\ket{a}$. The complexity of $U_{\rm FT}(\xi)$ is $O(\mathcal{K} \log \mathcal{K})$, and each of its gates must be controlled by $\log \mathcal{V}$ qubits. The resulting complexity of the controlled $U_{\rm FT}(\xi)$ is $O[(\log_2^2\mathcal{V})\mathcal{K}(\log_2 \mathcal{K})]$. There are $O(\mathcal{\mathcal{V}})$ such controlled circuits over all possible values of $\xi$. Therefore, the elementary-gate complexity of $U_{\Pi^{Ia}}$ is
\begin{align}
    \mathcal{M}\left( U_{\tilde{\Pi}^{I}} \right) &\sim O\left(\mathcal{V}\mathcal{K}\left[(\log_2^2\mathcal{V})(\log_2 \mathcal{K}) + \log_2^2(\mathcal{V}\mathcal{K})\right]\right)\nonumber\\\,
    & \sim O\left(\mathcal{V}\mathcal{K}\log_2^2(\mathcal{V}\mathcal{K})\right)\,.
\end{align}
where the first and second terms in the first line arise from the controlled $U_{\rm FT}$ and controlled $Z_{\mathcal{J}\,\xi}$ operations, respectively.

\subsection{Block encoding of composite operators}\label{sec:block_KQ}

The composite operators $K$, $K^{-1}$, $\ad(A_I)$, $D_I[A]$, and $Q$ can be block encoded using the elementary block encodings constructed above. Because $K$ and $K^{-1}$ act only on the position space, we discuss them first. We then construct block encodings of $\ad(A_I)$, $D_I[A]$, and $Q$.

\subsubsection{Block encoding of \texorpdfstring{\(K\)}{K} and \texorpdfstring{\(K^{-1}\)}{K-1}}\label{subsubsec:block_KA}

Recall that $K$ is
\begin{align}
    K &= K_3 + K_2+K_1 \nonumber\\
    &= (\partial_3^R)^\dagger\partial_3^R
    +\delta_{n_3,0} (\partial_2^R)^\dagger\partial_2^R + \delta_{n_2,0}\delta_{n_3,0} (\partial_1^R)^\dagger\partial_1^R\nonumber\\
    &= -\partial^2_3 - \delta_{n_3,0} \partial_2^2 - \delta_{n_2,0}\delta_{n_3,0} \partial_1^2\,,
\end{align}
where the last equality follows from $(\partial^R)^\dagger = -\partial^L$ and $\partial^2 = \partial^L \partial^R$.

To complete the block encoding of $K$, we also need block encodings of $\delta_{n_3,0}$ and $\delta_{n_2,0}$. In the space $\Spos$, $\delta_{n_3,0}$ is represented by
\begin{align}
    &\delta_{n_3,0} = \ket{00..0} \bra{00...0}_3 = \frac{1}{2}\left(I_3-R_3^{(0)}\right)\,,\nonumber\\
    &R_3^{(0)} = I_3 - 2\ket{00,...,0}\bra{00,...,0}_3\,,
\end{align}
The operator $R_3^{(0)}$ is obtained from the reflection operator in Eq.~\eqref{eq:PR3} by replacing $\ket{11...1}$ with $\ket{00...0}$. This replacement can be implemented by a layer of $X$ gates with complexity $O(\log_2\mathcal{V})$. We can then construct a $(1,1,0)$ block encoding of $\delta_{n_3,0}$, denoted by $U_{\delta_{n_3,0}}$. A $(1,1,0)$ block encoding $U_{\delta_{n_2,0}}$ of $\delta_{n_2,0}$ can be constructed in the same way.

The block encoding of $K_3$ is
\begin{align}
    U_{K_3} = (U^\dagger_{\partial_3})^{(r)} U^{(r')}_{\partial_3}
\end{align}
This is a $(4,4,0)$ block encoding of $K_3$. The superscripts $r$ and $r'$ differ because $U^{(r')}_{\partial_3}$ and $U^{(r)}_{\partial_3}$ act on distinct ancilla registers. Consequently, $(U^\dagger_{\partial_3})^{(r)} U^{(r')}_{\partial_3} \not = I$. Products of block-encoding operators always use distinct ancilla registers, so we suppress the superscripts $(r)$ and $(r')$ hereafter. Similarly,
\begin{align}
    U_{K_2} = U_{\delta_{n_3,0}} U^\dagger_{\partial_2} U_{\partial_2}\,,
\end{align}
is a $(4,5,0)$ block encoding of $K_2$, and
\begin{align}
    U_{K_1} = U_{\delta_{n_2,0}}U_{\delta_{n_3,0}} U^\dagger_{\partial_1} U_{\partial_1}\,,
\end{align}
is a $(4,6,0)$ block encoding of $K_1$. Finally, a $(\alpha_K = 12,a_K = 8,0)$ block encoding $U_K$ of $K$ can be constructed by combining the $U_{K_i}$ using LCU. The gate complexity $\mathcal{M}(U_K)$ is dominated by the cost of implementing $T_i$:
\begin{align}
    \mathcal{M}({U_K}) \sim \mathcal{M}({T_i}) \sim O(\log_2^3 \mathcal{V})\,.
\end{align}

A block encoding of $K^{-1}$, denoted by $U_{K^{-1}}$, can be constructed from $U_K$ using quantum singular value transformation (QSVT). Its complexity is determined by the condition number $\kappa \sim O(\mathcal{V}^{4/3})$ of $K$. According to Ref.~\cite{PRXQuantum.2.040203}, a $(\Gamma,8,\varepsilon_{\rm inv})$ block encoding $U_{K^{-1}}$ of $K^{-1}$ can be constructed using $d_{\rm inv}$ queries to $U_K$, where
\begin{align}
    d_{\rm inv} \sim O\left(\kappa \log(\kappa/\varepsilon_{\rm inv})\right)\,.
\end{align}
The resulting gate cost for implementing $K^{-1}$ is
\begin{align}
    \mathcal{M}(U_{K^{-1}}) \sim  O\left(\mathcal{M}(U_K) \kappa \log(\kappa/\varepsilon_{\rm inv})\right) \sim O\left(\mathcal{V}^{4/3} (\log_2^3\mathcal{V})\log(\mathcal{V}\varepsilon_{\rm inv}^{-1})\right)\,.
\end{align}

\subsubsection{Block encoding of \texorpdfstring{\(\ad(A_I)\)}{adA}, \texorpdfstring{\(D^L_I[A]\)}{DI}, \texorpdfstring{\(\mathcal{Q}\)}{Q}, and \texorpdfstring{\(F^a_{ij}\)}{F}}
The operator $\ad(A_I)$ appears in both $\mathcal{Q}^a(n)$ and $F^a_{IJ}(n)$. In the space $\Spos\otimes\Scol\otimes \mathcal{H}$, it is given by
\begin{align}
    \ad(A_I)
    =
    \sum_{n,a,b,c}
    f^{abc}\ket{n}\ket{a}\!\bra{n}\bra{c}
    \otimes A_I^b(n).
    \label{eq:v4-adA-indexed}
\end{align}
Define a set of matrices acting on the color space $\Scol$ by
\begin{align}
    M^b := \sum_{ac} f^{abc} \ket{a} \bra{c}
    \label{eq:v4-Mb-def}
\end{align}
The operator $\ad(A_I)$ can then be decomposed as
\begin{align}
    \ad(A_I)
    =
    \sum_{b=1}^{N_{\rm adj}}
    M^b
    \left[
    \sum_{n}
    \ket{n}\!\bra{n}
    A_I^b(n)
    \right] = \sum_{b} M^b \bar{A}^b_I(n)\,.
    \label{eq:v4-adA-Mb-decomp}
\end{align}
The operator $U_{\bar{A}^a_I}$ block encodes $\sum_{n} \ket{n}\bra{n} A^a_I(n)$, so it remains to block encode the $N_{\rm adj} \times N_{\rm adj}$ matrix $M^b$. Choose $\alpha_M \geq ||M^b||$ for every $b$ and define $\bar{M}^b = M^b/\alpha_M$. A block encoding of $M_b$ can then be constructed by introducing one ancilla qubit $\ket{0}_M$:
\begin{align}
    U_{M^b}
    =
    \begin{pmatrix}
    \bar M^b
    &
    \sqrt{I-\bar M^b\bar M^{b\dagger}}
    \\
    \sqrt{I-\bar M^{b\dagger}\bar M^b}
    &
    -\bar M^{b\dagger}
    \end{pmatrix}\,,
    \label{eq:v4-Mb-dilation}
\end{align}
which is a \((\alpha_M,1,0)\)-block encoding of
\(M^b\):
\begin{align}
    M^b
    =
    \alpha_M
    \bra 0_M U_{M^b}\ket 0_M.
    \label{eq:v4-Mb-BE}
\end{align}

We can now construct $\ad(A_I)$ using $U_{M^b}$, $U_{\bar{A}_I^a}$, and a $\log_2 N_{\rm adj}$-qubit ancilla register initialized in $\ket{0^{\log_2N_{\rm adj}}}$. The ${\rm PREP}_{\rm adj}$ gate is defined by
\begin{align}
    {\rm PREP}_{\rm adj} \ket{0^{\log_2 N_{\rm adj}}} = \sum_{a=0}^{N_{\rm adj} -1} \frac{1}{\sqrt{N_{\rm adj}}} \ket{a}\,,
\end{align}
which is implemented by a layer of $H$ gates acting on the ${\rm adj}$ register. The ${\rm SELECT}_{\rm adj}$ gate is
\begin{align}
    {\rm SELECT}_{\rm adj} = \sum_{a} \ket{a}\bra{a} U_{M^a} U_{\bar{A}^a_I}\,.
\end{align}
This gate can be constructed from controlled versions of $U_{M^a}$ and $U_{\bar{A}^a_I}$. We then obtain the $(\alpha_{\ad(A_I)},a_{\ad(A_I)},0)$ block encoding of $\ad(A_I)$:
\begin{align}
    U_{\ad(A_I)} = {\rm PREP}_{\rm adj}^\dagger {\rm SELECT}_{\rm adj} {\rm PREP}_{\rm adj}\,,
\end{align}
where
\begin{align}
    &\alpha_{\ad} = N_{\rm adj} \alpha_M A_{\rm max} \sim O(A_{\rm max})\,,\nonumber\\
    &a_{\ad} \sim O(\log_2\mathcal{K}+\log_2 N_{\rm adj})\sim O(\log_2\mathcal{K})\,.
\end{align}
The elementary-gate complexity of $U_{\ad(A_I)}$ has the same scaling as $\mathcal{M}(U_{\bar{A}^b_{I}})$:
\begin{align}
    \mathcal{M}\left(U_{\ad({A}_I)}\right) \sim \mathcal{M}\left(U_{\bar{A}^b_{I}}\right) \sim O\left(\mathcal{V}\mathcal{K}\log_2^2 (\mathcal{V}\mathcal{K})\right)\,.
\end{align}

The covariant derivative $D^L_I[A] = \partial^L_I + g\ad(A_I)$ can also be block encoded by LCU because $U_{\partial_I}^\dagger$ and $U_{\ad(A_I)}$ have already been constructed. Set
\begin{align}
    \alpha_{D}
    &=\alpha_{\partial} + g\alpha_{\ad} \sim O\left(gA_{\rm max}\right).
    \label{eq:v4-alpha-D}
\end{align}
The number of ancilla qubits required to block encode $D^L_I$ scales as
\begin{align}
    a_{D} \sim a_{\ad} \sim O(\log_2\mathcal{K})\,.
\end{align}
This yields an $(\alpha_{D},a_D,0)$ block encoding of $D^L_I[A]$, denoted by $U_{\bar{D}_I}$. Projecting all ancilla qubits onto $\ket{0^{a_D}}$ gives
\begin{align}
    \bra{0^{a_D}} U_{\bar{D}_I}\ket{0^{a_D}} = \frac{1}{\alpha_\partial + g\alpha_{\ad}} \left(\partial^L_I + g\sum_{n,a,b,c} \ket{n}\ket{a} \bra{n}\bra{c} f^{abc} A_I^b(n)\right)\,. 
\end{align}

We now construct a block encoding of the charge-density operator $\mathcal{Q}^a$. The operator of interest is
\begin{align}
    \tilde{\mathcal{Q}} = \sum_{n,a} \ket{n}\ket{a} {\mathcal{Q}}^a(n)\,.
\end{align}
Using $\mathcal{Q}(n) = \partial^L_I \Pi^{I} + g\ad(A_I)\Pi^I = D^L_I \Pi^I$, we obtain a block encoding of $\tilde{\mathcal{Q}}$ by combining $U_{\bar{D}_1}U_{\tilde{\Pi}_1}$ and $U_{\bar{D}_2}U_{\tilde{\Pi}_2}$ through LCU. As an example, consider
\begin{align}
    \bra{0^{a_D,a_{\tilde{\Pi}}}} U_{\bar{D}_1} U_{\tilde{\Pi}_1} \ket{0^{a_D,a_{\tilde{\Pi}}}}\ket{0} &= \frac{1}{\alpha_\partial + g\alpha_{\ad}} \frac{1}{\sqrt{\mathcal{V} N_{\rm adj}}\Pi_{\rm max}} \left(\partial_1^L + \sum_{n,a,b,c} \ket{n}\ket{a}\bra{n}\bra{c} g f^{abc} A^b_1(n)\right)\\
    &\quad \times \left(-\sum_{m,d} \ket{m}\ket{d} \Pi^{d}_1(m) \right)\nonumber\\
    &= \frac{1}{\alpha_\partial + g\alpha_{\ad}} \frac{1}{\sqrt{\mathcal{V} N_{\rm adj}}\Pi_{\rm max}} \sum_{n,a} \ket{n}\ket{a}\left( -\partial_1 \Pi^{a}_1(n) - g f^{abc} A_1^b(n) \Pi^{c}_1(n)\right)\,,
\end{align}
where $\ket{0}$ is an initial state in $\Spos\otimes\Scol$. Denoting the block encoding of $\tilde{\mathcal{Q}}$ by $U_{\tilde{\mathcal{Q}}}$, projection of the ancilla qubits onto $\ket{0}$ gives
\begin{align}
    \bra{0^{a_{\tilde{\mathcal{Q}}}}} U_{\tilde{\mathcal{Q}}} \ket{0^{a_{\tilde{\mathcal{Q}}}}}\ket{0} = \sum_{n,a} \frac{1}{2} \frac{1}{\alpha_\partial + g \alpha_{\ad}} \frac{1}{\sqrt{\mathcal{V}N_{\rm adj}}\Pi_{\rm max}} \ket{n}\ket{a} \mathcal{Q}^a(n)\,.
    \end{align}
Thus, $U_{\tilde{\mathcal{Q}}}$ is an $(\alpha_{\tilde{\mathcal{Q}}},a_{\tilde{\mathcal{Q}}},0)$ block encoding of $\tilde{\mathcal{Q}}$, with
\begin{align}
    &\alpha_{\tilde{\mathcal{Q}}} = 2\sqrt{\mathcal{V} N_{\rm adj}} \Pi_{\rm max}(\alpha_\partial + g \alpha_{\ad}) \sim O(g\mathcal{V}^{1/2} \Pi_{\rm max} A_{\rm max})\,,\nonumber\\
    &a_{\tilde{\mathcal{Q}}} \sim O(\log_2\mathcal{K})\,.
\end{align}
The gate complexity of $U_{\tilde{\mathcal{Q}}}$ is
\begin{align}
    \mathcal{M}(U_{\tilde{\mathcal{Q}}}) \sim \mathcal{M}(U_{\tilde{\Pi}^I}) \sim  O\left(\mathcal{V}\mathcal{K}\log_2^2(\mathcal{V}\mathcal{K})\right)\,.
\end{align}

Finally, we consider the block encoding of
\begin{align}
    \tilde{F}_{ij} &= \sum_{n,a} \ket{n}\ket{a} F^a_{ij}(n)\,.
\end{align}
Equation~\eqref{eq:defFIJ} shows that a block encoding of $\tilde{F}_{ij}$, denoted by $U_{\tilde{F}_{ij}}$, can be constructed using $U_{\partial_i}$, $U_{\tilde{A}_J}$, and $U_{\ad(A_I)}$. For example, the $\sum_{n,a}\ket{n}\ket{a}\partial_2 A_1^a(n)$ contribution to $\tilde{F}_{21}$ is
\begin{align}
    \bra{0^{a_\partial,a_A}}U_{\partial_2} U_{\tilde{A}_1}\ket{0^{a_\partial,a_A}} \ket{0} = \frac{1}{2\sqrt{\mathcal{V}N_{\rm adj}}A_{\rm max}} \sum_{n,a}\ket{n}\ket{a} \partial^R_2 A_1^a(n)\,.
\end{align}
The $\ad(A_2)A_1$ contribution to $\tilde{F}_{21}$ is
\begin{align}
    \bra{0^{a_{\ad},a_A}}U_{\ad(A_2)} U_{\tilde{A}_1}\ket{0^{a_{\ad},a_A}}\ket{0} = \frac{1}{\alpha_M\sqrt{\mathcal{V}N_{\rm adj}^3}A_{\rm max}^2} \sum_{n,a}\ket{n}\ket{a} f^{abc} A^b_2(n) A^c_1(n)\,.
\end{align}
Thus, $\tilde{F}_{21}$ can be obtained by combining $U_{\partial_2}U_{\tilde{A}_1}$, $U_{\partial_1}U_{\tilde{A}_2}$, and $U_{\ad(A_2)} U_{\tilde{A}_1}$ using LCU. Because gauge fixing imposes $A_3^a = 0$, the $F_{32}$ and $F_{31}$ terms require only $U_{\partial_3}U_{\tilde{A}_2}$ and $U_{\partial_3}U_{\tilde{A}_1}$, respectively. We denote the $(\alpha_{\tilde{F}},a_{\tilde{F}},0)$ block encoding of $\tilde{F}_{ij}$ by $U_{\tilde{F}_{ij}}$, where
\begin{align}
    &\alpha_F = 4\sqrt{\mathcal{V}N_{\rm adj}}A_{\rm max} + g\alpha_M\sqrt{\mathcal{V}N_{\rm adj}^3}A_{\rm max}^2 \sim O(g\mathcal{V}^{1/2} A_{\rm max}^2)\,,\nonumber\\
    &a_F \sim O(\log_2 \mathcal{K})\,.
\end{align}
Here, we use the value of $\alpha_F$ for $U_{\tilde{F}_{12}}$ as the common normalization for all $F_{ij}$, because the operator-norm bound for $\tilde{F}_{12}$ is the largest among those for the $\tilde{F}_{ij}$. The implementation cost of $U_{\tilde{F}_{ij}}$ is dominated by that of $U_{\ad(A_I)}$:
\begin{align}
    \mathcal{M}(U_{\tilde{F}_{ij}}) \sim \mathcal{M}(U_{\ad(A_I)}) \sim O\left(\mathcal{V}\mathcal{K}\log_2^2 (\mathcal{V}\mathcal{K})\right)\,.
\end{align}

\subsection{Block encoding of $\Hfix$ and estimation of $\|H_{\rm fix}\|$}\label{sec:blHfix}
We now construct the block encoding of $\Hfix$ and obtain a corresponding bound on $\|H_{\rm fix}\|$. We begin with
\begin{align}
    H_\Pi = \frac{1}{2}\sum_n (\Pi_I^a(n))^2\,.
\end{align}
This term can be block encoded using $U^\dagger_{\tilde{\Pi}_I}U_{\tilde{\Pi}_I}$. Projecting the ancilla qubits and the $\Spos\otimes\Scol$ register onto $\ket{0}$ gives
\begin{align}
    \bra{0}\bra{0^{a_{\tilde{\Pi}},a'_{\tilde{\Pi}}}} U^{\dagger}_{\tilde{\Pi}_2} U_{\tilde{\Pi}_2} \ket{0^{a_{\tilde{\Pi}},a'_{\tilde{\Pi}}}} \ket{0} = \frac{1}{\mathcal{V}N_{\rm adj}\Pi_{\rm max}^2} \sum_{n} [\Pi^a_2(n)]^2\,,
\end{align}
where the $\Pi_2^a$ term is used as an example. Combining $U^{\dagger}_{\tilde{\Pi}_1}U_{\tilde{\Pi}_1}$ and $U^{\dagger}_{\tilde{\Pi}_2}U_{\tilde{\Pi}_2}$ using LCU yields an $(\alpha_{H_\Pi},a_{H_\Pi},0)$ block encoding of $H_\Pi$, denoted by $U_{H_\Pi}$, with
\begin{align}
    &\alpha_{H_\Pi} = \mathcal{V} N_{\rm adj} \Pi_{\rm max}^2 \sim O(\mathcal{V} \Pi_{\rm max}^2)\,,\nonumber\\
    &a_{H_\Pi} \sim O(\log_2\mathcal{K} + \log_2\mathcal{V})\,.
\end{align}
The complexity of $U_{H_\Pi}$ is determined by $\mathcal{M}(U_{\tilde{\Pi}^I})$; hence,
\begin{align}
    \mathcal{M}(U_{H_\Pi}) \sim O\left(\mathcal{V}\mathcal{K}\log_2^2(\mathcal{V}\mathcal{K})\right)\,.
\end{align}

We next consider the block encoding of
\begin{align}
    H_B = \frac{1}{2} \sum_n \sum_{i>j} \left[F^a_{ij}(n)\right]^2\,.
\end{align}
The $\sum_n[F^a_{21}(n)]^2$ term in $H_B$ can be implemented using $U^{\dagger}_{\tilde{F}_{21}}U_{\tilde{F}_{21}}$:
\begin{align}
    \bra{0}\bra{0^{a_F,a'_F}} U^{\dagger}_{\tilde{F}_{21}}U_{\tilde{F}_{21}} \ket{0^{a_F,a'_F}}\ket{0} = \frac{1}{\alpha_F^2} \sum_n [F^a_{21}(n)]^2\,.
\end{align}
Similarly, $U^{\dagger}_{\tilde{F}_{31}} U_{\tilde{F}_{31}}$ and $U^{\dagger}_{\tilde{F}_{32}} U_{\tilde{F}_{32}}$ block encode $\sum_{n}[F^a_{31}(n)]^2$ and $\sum_n[F^a_{32}(n)]^2$, respectively. Combining these three block encodings of $F^2$ using LCU yields an $(\alpha_{H_B},a_{H_B},0)$ block encoding of $H_B$, denoted by $U_{H_B}$, with
\begin{align}
    &\alpha_{H_B} = \frac{3\alpha_F^2}{2} \sim O\left(g^2 \mathcal{V} A_{\rm max}^4\right)\,,\nonumber\\
    &a_{H_B} \sim O(\log_2\mathcal{V} + \log_2\mathcal{K})\,.
\end{align}
The gate complexity of $U_{H_B}$ is determined by that of $U_{\tilde{F}_{ij}}$; hence,
\begin{align}
    \mathcal{M}(U_{H_B}) \sim \mathcal{M}(U_{\tilde{F}_{ij}}) \sim O\left(\mathcal{V}\mathcal{K}\log_2^2 (\mathcal{V}\mathcal{K})\right)\,.
\end{align}

Finally, consider
\begin{align}
    H_V = \frac{1}{2}\sum_{n,m} [\mathcal{Q}^a(n)]^\dagger K^{-1}_{n,m} \mathcal{Q}^a(m)\,.
\end{align}
A block encoding of $H_V$ can be constructed from $U_{\tilde{Q}}$ and $U_{K^{-1}}$:
\begin{align}
    \bra{0}\bra{0^{a_{\tilde{\mathcal{Q}}},a_K,a'_{\tilde{\mathcal{Q}}}}} U^{\dagger}_{\tilde{\mathcal{Q}}} U_{K^{-1}} U_{\tilde{\mathcal{Q}}} \ket{0^{a_{\tilde{\mathcal{Q}}},a_K,a'_{\tilde{\mathcal{Q}}}}} \ket{0} = \frac{2}{\alpha_{\tilde{Q}}^2 \Gamma} H_V\,.
\end{align}
Thus, $U^{\dagger}_{\tilde{\mathcal{Q}}} U_{K^{-1}} U_{\tilde{\mathcal{Q}}}$ is an $(\alpha_{H_{V}},a_{H_V},\varepsilon_{H_V})$ block encoding of $H_V$, with
\begin{align}
    &\alpha_{H_V} = \frac{\alpha_{\tilde{Q}}^2 \Gamma}{2} \sim O\left(g^2 \mathcal{V}^{7/3} \Pi_{\rm max}^2 A_{\rm max}^2\right)\,,\nonumber\\
    &a_{H_V} \sim O\left(\log_2(\mathcal{K}\mathcal{V})\right)\,,\nonumber\\
    &\varepsilon_{H_V} = \alpha_{\tilde{\mathcal{Q}}}^2 \varepsilon_{\rm inv} \sim O\left(g^2 \mathcal{V} \Pi_{\rm max}^2 A_{\rm max}^2 \varepsilon_{\rm inv}\right)\,.
\end{align}
From the main text, we have $\Pi_{\rm max}^2 A_{\rm max}^2 \sim O(\mathcal{V}^3 E^2/\varepsilon_s^2)$, thus
\begin{align}
    \alpha_{H_V} \sim O\left(\frac{\mathcal{V}^{16/3}E^2g^2}{\varepsilon_s^2}\right)
\end{align}
The gate complexity of $U_{H_V}$ is dominated by that of $U_{K^{-1}}$; hence,
\begin{align}
    \mathcal{M}({U_{H_V}}) \sim \mathcal{M}(U_{K^{-1}}) \sim O\left(\mathcal{V}^{4/3} (\log_2^3\mathcal{V})\log(\mathcal{V}\varepsilon_{\rm inv}^{-1})\right)\,.
\end{align}

A block encoding of $\Hfix$ can be constructed by combining $U_{H_\Pi}$, $U_{H_B}$, and $U_{H_V}$ using LCU. This yields an $(\alpha_{\Hfix},a_{\Hfix},\varepsilon_{\Hfix})$ block encoding of $\Hfix$, denoted by $U_{\Hfix}$, with
\begin{align}
    &\alpha_{\Hfix} \sim \alpha_{H_V} \sim O\left(\frac{\mathcal{V}^{16/3}E^2g^2}{\varepsilon_s^2}\right)\,,\nonumber\\
    &a_{\Hfix} \sim \log_2(\mathcal{V}\mathcal{K})\,,\nonumber\\
    &\varepsilon_{\Hfix} = \varepsilon_{H_V} \sim O\left(\frac{\mathcal{V}^4 E^2 g^2 \varepsilon_{\rm inv}}{\varepsilon_s^2}\right)\,.
\end{align}
The resulting gate complexity of $U_{\Hfix}$ is
\begin{align}
    \mathcal{M}(U_{\Hfix}) \sim \mathcal{M}(U_{H_V}) \sim O\left(\mathcal{V}^{4/3}(\log_2^3\mathcal{V})\log(\mathcal{V}\varepsilon_{\rm inv}^{-1})\right)\,.
\end{align}

\SMfinish

\end{document}